\documentclass[review]{elsarticle}
\makeatletter \def\ps@pprintTitle{  \let\@oddhead\@empty  \let\@evenhead\@empty  \def\@oddfoot{\hfill\thepage}  \def\@evenfoot{\thepage\hfill}} \makeatother
\usepackage{lineno}

\usepackage{url} 
\usepackage{amssymb}
\usepackage{rotating,tabularx}
\usepackage[labelsep=period]{caption}
\usepackage{subcaption}
\usepackage{epstopdf}
\usepackage{amsmath}
\usepackage{amsthm}
\usepackage{multirow}
\usepackage{mathtools}
\usepackage{makecell}
\usepackage{booktabs}
\usepackage{color}
\usepackage{float}
\usepackage{setspace}
\usepackage{xspace}
\usepackage{comment}
\usepackage[colorlinks=true,linkcolor=blue,urlcolor=blue,citecolor=blue]{hyperref}
\setcitestyle{notesep={; },round,aysep={},yysep={;}}
\usepackage[utf8]{inputenc}
\usepackage[english]{babel}

\usepackage{epstopdf}
\usepackage{etoolbox}

\usepackage{cleveref}
\usepackage{svg}
\usepackage{graphicx}
\usepackage{enumerate}
\usepackage{algorithm}
\usepackage{algpseudocode}
\usepackage{url} 
\usepackage[titletoc,toc,title]{appendix}
\usepackage[dotinlabels]{titletoc}

\usepackage{array}

\newcolumntype{L}[1]{>{\raggedright\let\newline\\\arraybackslash\hspace{0pt}}m{#1}}
\newcolumntype{C}[1]{>{\centering\let\newline\\\arraybackslash\hspace{0pt}}m{#1}}
\newcolumntype{R}[1]{>{\raggedleft\let\newline\\\arraybackslash\hspace{0pt}}m{#1}}

\newcommand{\citcount}{\#Cit\xspace}
\biboptions{authoryear}

\begin{document}

\begin{frontmatter}
\title{Article's Scientific Prestige: measuring the impact of individual articles in the Web of Science}





\author[a,b]{Ying Chen}
\author[c,d]{Thorsten Koch}
\cortext[mycorrespondingauthor]{Corresponding author}
\ead{koch@zib.de}
\author[d]{Nazgul Zakiyeva}
\author[e]{\\Kailiang Liu}
\author[f]{Zhitong Xu}
\author[g]{Chun-houh Chen}
\author[h]{Junji Nakano}
\author[i]{\\Keisuke Honda}

\address[a]{Department of Mathematics, National University of Singapore, Block S17, Level 4, 2 Science Drive 2, 117543 Singapore}
\address[b]{Risk Management Institute,  National University of Singapore, 21 Heng Mui Keng Terrace, 04-03, 119613 Singapore}
\address[c]{Technische Universit{\"a}t Berlin, Chair of Software and Algorithms for Discrete Optimization, Stra{\ss}e des 17. Juni 135, 10623 Berlin, Germany
}
\address[d]{Zuse Institute Berlin, Applied Algorithmic Intelligence Methods Department, Takustra{\ss}e 7, 14195 Berlin, Germany
}
\address[e]{National University of Singapore (Chongqing) Research Institute, Building 5B, Chongqing Internet Institute, No. 16 South Huashan Road, Chongqing Liang Jiang New Area, Chongqing, China
}
\address[f]{Department of Statistics and Data Science, National University of Singapore, Block S16, Level 7, 6 Science Drive 2, 117546
Singapore}
\address[g]{The Institute of Statistical Science, Academia Sinica,  128 Academia Road, Section 2, Nankang, 11529 Taipei,  Taiwan
}
\address[h]{Department of Global Management, Chuo University, 742-1 Higashinakano, Hachioji, 192-0393 Tokyo,  Japan}
\address[i]{Institute of Statistical Mathematics, 10-3 Midorichō, Tachikawa, 190-0014 Tokyo, Japan }

\begin{abstract}
We performed a citation analysis on the Web of Science publications consisting of more than 63 million articles and 1.45 billion citations on 254 subjects from 1981 to 2020. We proposed the Article’s Scientific Prestige (ASP) metric and compared this metric to number of citations (\citcount) and journal grade in measuring the scientific impact of individual articles in the large-scale hierarchical and multi-disciplined citation network. In contrast to \citcount, ASP, that is computed based on the eigenvector centrality, considers both direct and indirect citations, and provides steady-state evaluation cross different disciplines. 
We found that ASP and \citcount are not aligned for most articles, with a growing mismatch amongst the less cited articles. 
While both metrics are reliable
for evaluating the prestige of articles such as Nobel Prize winning articles, ASP tends to provide more persuasive rankings than
\citcount when the articles are not highly cited. The journal grade, that is eventually determined by a few highly cited articles, is unable to properly reflect the scientific impact of individual articles. 
The number of references and coauthors are less relevant to scientific impact, but subjects do make a difference.
\end{abstract}


\end{frontmatter}


\section{Introduction}
Known as the most popular deliverable of scientific research, the peer reviewed article is considered a main carrier of new knowledge and information, presenting innovative findings, demonstrating unique contributions, and promoting openness and transparency in science. It is apparent that individual articles have different scientific impacts. Given the ever growing number of publications in science, quantifying an article’s scientific prestige has been an important topic to fairly evaluate its contribution to the scientific progress, see, for example, \cite{chu2021slowed,chang2019new, zhao2022utilizing, li2019deep, nie2019academic, tahamtan2016factors, xiao2016modeling}. 

Given that all parts of science compete on the available research funding resources, and that universities, and even countries try to evaluate and compare their respective scientific impact, we consider the problem of measuring the scientific prestige of individual articles in a setting where the size of the citation network is huge — that is, where the number of nodes (articles) and edges (citations/references) is at the million/billion level and all the disciplines in science are considered. 
We propose the Article’s Scientific Prestige (ASP) metric, based on the recent advances in eigenvector centrality (or Pagerank) and optimization to address this large-scale data analysis challenge with computational tractability. More importantly, we attempt to perform a comprehensive citation analysis of all the published articles in various disciplines and over time, and provide a scientific comparison of several citation metrics at the level of individual articles.

Our approach is motivated by a specific application: measuring the scientific impact of each individual article in the Web of Science (WoS) citation network. The top influential papers are easy to spot. They introduce new terms and names and initiate research in a new area. The least influential papers are also easy to identify, as those articles are never cited and thus have negligible impact.
Evaluating the remaining articles is however challenging and how this should be done remains an open question. The Number of citations (\citcount) and the journal grade have long been used as metrics to show how much attention an article has received in the science community. Note that the popular metrics are counting the number of times an entity (e.g., a scholar, an institute, or a journal), rather an article, has been cited, see, e.g., the Science Citation Index (SCI) by \cite{garfield1955citation}, CiteScore by \cite{garfield1972citation}, H-index by \cite{hirsch2005index}, and the SCImago Journal Rank (SJR) by \cite{gonzalez2010new}.
It is hypothesised that the more citations an entity obtains, the higher scientific impact it has. Statistically speaking, such an aggregation lowers the randomness and misjudge chance for an entity with many publications compared to an individual article. Simultaneously, it trivializes the individual impact of each scientific work. Alternatively, it becomes common and recognized to judge an article’s scientific value by journal grade, i,e. which journal the article is published. Scholars partition journals in classes like A*, A, B, C, and imply that at least on average the grade of the journal reflects the quality of the articles it publishes.  

Our main question of interest is to what extent the \citcount and journal grade are helpful to assess the impact of an individual article given a citation network. Though popularly adopted in all kinds of evaluations, one can easily find counter examples where either way fails. It has been acknowledged that \citcount, though direct and convenient, is not comparable across disciplines and over time given different publication frequency and citation duration. It has also been argued that self-citation (i.e., author cites their own articles in another article) or community-citation (different authors, yet with strong academic connections) can easily abuse the metric. As for journal grade, a large portion of the articles have a much lower impact than the journal's average given the extremely skewed distribution of citations. Even the top-tiered journals have a substantial number of articles that are not cited at all, implying one should not judge an article (solely) based on which journal publishes it. 

We perform a large-scale analysis on the Web of Science (WoS) citation network, with more than 63 million articles and 1.45 billion citations on 254 subjects from 1981-2020\footnote{After removing self-citations and articles without references or without a subject, we look at 33,200,017 articles with 898,879,235 references from 254 subjects. There are 255 subjects in WoS. However, articles in the subject ``Planning and development" do not have references.}. To  demonstrate the spectrum of citations on various disciplines, we compute the ASP of all the articles together, with which we assess the scientific influence of individual articles in the network. To obtain an accurate quantitative measure of scientific prestige, we must solve two technical challenges. The first is the large scale of the citation network, which requires an efficient optimization approach with computational tractability. Second is the hyperparameter choice in the eigenvector centrality computation such as the damping factor, ensuring a stable performance and also fair comparisons among various disciplines over time.

We implement a parallel Jacobi iterations based algorithm on sparse data-structures
to compute a steady-state solution for the ASP values, see \cite{GolubVanLoan2013} and \cite{srivastava2019aitken}.
Running on an 8-core Intel Core i7-9700K CPU at 3.60GHz, the algorithm takes less than 2 seconds per
iteration and converges in less than 20 iterations. The efficient algorithm allows for a wide range search of hyperparameters, even for large-scale citation networks. Specifically, we determine the damping parameter to $0.5$ and also adopt a citing window of 5 years for the optimal stability of scientific contributions over disciplines and time.

We found that ASP and \citcount are not aligned for most articles, with a growing mismatch amongst the less cited articles, although the two metrics display similar ranks among the top 10\% highly cited articles and are identical for the bottom 20-30\% of articles (as those are never cited). The journal grade, that is eventually determined by a few highly cited articles, is unable to properly reflect the scientific impact of individual articles. When aggregating to the journal level, ASP is more consistent with the journal grade than \citcount. Moreover, we found that articles with the largest ASP and \citcount were in the subjects of Science, Biology, and Geography, and the smallest in Social Science, Arts, Law \& Policy, and Education. The number of references and coauthors are less relevant to scientific impact, but subjects do make a difference.

We build our analysis on pioneering works. Many aspects of the current work, including data (the size, time interval, and diversity), algorithm (to estimate the eigenvector centrality metric), and the empirical investigations at article level are novel with respect to the prior works. \cite{massucci2019measuring} considered the citation network dataset of 5 disciplines -- Dentistry, Oral Surgery \& Medicine; Business \& Finance; Information Science \& Library Science; Telecommunication; and Veterinary Sciences -- from 2010 to 2014 provided by Clarivate Analytics, and analyzed citation patterns at a university level. \cite{ma2008bringing} studied 236,517 articles in Biochemistry and Molecular Biology from 2003 to 2005 based on the Institute for Scientific Information (ISI) database, see also \cite{palacios2004measurement}. In terms of data size, \cite{chu2021slowed} conducted also a large-scale citation analysis with WoS data from 1960 to 2014. The focus is to show that the gigantic increase of articles may impede the rise of new ideas instead of promoting the rate of scientific progress. Our paper is also related to other works on eigenvector centrality or Pagerank based metrics. \cite{bergstrom2007eigenfactor} and \cite{gonzalez2010new} computed a citation metric for academic journal evaluations (i.e., at journal level) and the latter demonstrated the application on articles from 296 subjects but published in year 2007 only with the Scopus database.

Our paper contributes a multi-disciplined citation analysis via connectivity in an extensive citation network. For using eigenvector centrality, we measure the influence of individual articles and provide a scientific comparison and statistics summary of several citation metrics at the level of individual article. The framework that we develop can be applied to a broad class of citation analysis problems in which the goal can be quantify the impact of an entity in a high-dimensional setting. Meanwhile, we are limited to the references within our database. By incorporating articles from online platforms such as \textit{arXiv.org} and \textit{Social Science Research Network (SSRN)}, or data from \textit{crossref.org}, we can update the citation analysis in the future. 

The paper is organized as follows. Section \ref{sec:data} presents the Web of Science data. Section \ref{sec:method} details the method and the implementation algorithm. Section \ref{sec:asp} implements the ASP to evaluate the scientific contribution of articles. Section \ref{sec:asp_ncit} discusses the comparison of ASP with respect to \citcount and journal grade, as well as relation to coauthors and references. Section \ref{sec:conc} draws a conclusion. 

\section{Web of Science Data}
\label{sec:data}
Our primary source is the citation data of the Web of Science (WoS). We obtained the digital data from Clarivate Analytics via the Institute of Statistical Mathematics, Japan. WoS is an internet search platform that provides comprehensive citation data for 254 academic disciplines, including Natural Science, Technology, Social Sciences, Humanities, Arts, and so on. The WoS citation data contains 63,092,643 unique articles published in 65,045 journals\footnote{The average number of articles per journal per year is 94. The maximum number was 31,273 articles published in PLOSone in 2013.} with 953,967,411 citations over 40 years from 1981 to 2020. Each article contains a number of attributes, including information on the article (UID, Document type, DOI, Language, Title, Abstract, Discipline), author (Name and Affiliation), journal (Publisher Name, Journal Name, Year, Issue, Volume, Pages), and a list of references (citations received after publication and references cited in the article). 
See Appendix \ref{app1} for a sample observation of an article titled “Basic local alignment search tool” by Stephen Frank Altschul, Gish Warren, and others, published in 1990 in the \textit{Journal of Molecular Biology} which received 58,002 citations, the highest in the data. 

The input of our main analysis is a directed hierarchical graph, where each node (vertex) represents an article and each arc (link/arrow) represents a reference/citation. In contrast to our expectations, the graph resulting from the data was not an direct acyclic graph (DAG) implied by the toplogical order. The reason for this is that articles in the same year might reference each other, or, due to the different delays in review and publication, an article may reference a future article, leading to directed cycles in the graph.
We made the tree unidirectional in time by only including references to articles of the same year or before the publication date of the referencing article.
Before creating the citation tree, we performed the following pre-processing steps: 1) Remove 54,178 articles without subject information and 12,364,174 articles without any references; 2)  Ignore 346,988,766 reference links to publications outside of the data; 3) Restrict the analysis to articles published in the time frame of 1990 to 2015 only and count citations up to 5 years after publication of the article, but their references still traced back to 1981 and citations up to 2020 in the computation. The resulting citation network contained 33,200,017 articles with 898,879,235 references and 757,630,741 citations.

We restricted the analysis time period of articles to 1990-2015 to avoid boundary bias due to incomplete citations/references. The boundary effect is particularly severe in the earlier years such as 1981, where references are completely missing leading to broken edges, and more recently, where articles published in, for example, 2020 are cited less often than those published in, for example, 1990. We also chose 5 years as the citing window. Intuitively, a fixed window size standardizes the time frame of citation metrics and allows the comparison of scientific contributions fair between articles published a long time ago and those published recently. It also means one focuses on the relevant immediate scientific impact of an article over the 5 years after its publication. We admit that the 5-year window may not favour certain types of articles or disciplines such as pure theoretical articles or arts works which usually need a longer time to exhibit impact on science. 
However, this choice is justified by the quantitative parameter choice, see Section \ref{Sec:damping_factor}. Data seems to support the choice too given that the average age for an article receiving its first citation is 2.3 years. The choice of 5 years is occasionally consistent with the common evaluation period adopted by many academic entities.   

\begin{table}[ht!]
	\centering
	\caption{Statistical summary for \citcount,  References, and Coauthors for articles between 1990 and 2015.}
	\begin{tabular}{ l|r|r|r|r|r } 
		\toprule
		& Q1 & Median & Mean & Q3  & Max  \\
		\midrule
		\citcount   &	1 & 	7 &	22.82& 22 & 58,002\\
		\midrule
		\text{References}  & 10 &	 21 & 27.07  &	36 &	7,303
		 \\ 
		\midrule
		\text{(Co-)authors} &	2 &	3 & 4.13 & 5 &	5,576 \\
		\bottomrule
	\end{tabular}
	\label{tab:my_label1}
\end{table}

Table \ref{tab:my_label1} presents statistics of \citcount, References, and Coauthors of the WoS data from 1990 to 2015 at article level. In general, all features are right skewed distributed. The median of \citcount is 7 per article meaning that 50\% of the articles receive 7 or fewer citations within 5 years after publication, while the mean is almost triple this with a value of 22.82. The skewness is caused due to extremes. There are 22.53\% of the articles not cited at all. In contrast, 25\% of articles (Q3) have more than 22 citations, and the maximum citation counts reaches up to 58,002.  Analogously, the distributions of the number of references and coauthors are right skewed too but with less extreme values. About 75\% of articles contain 36 references or less. The maximum references is 7,303, in the article entitled ``Calcium-binding proteins 1: EF-hands” by Hiroshi Kawasaki and Robert H. Kretsinger (1995) published in \textit{Protein Profile}. There is an asymmetry between the References and \citcount. The median reference count of $21$ per article is triple to the median \citcount. This possibly implies ``citation clustering” where a few number of articles are commonly cited by many other articles. The number of coauthors remains low, with a median of 3, and 75\% of articles are written by less than 5 authors. Although, in Physics collaborations such as Atlas (Switzerland) and Compact Muon Solenoid, published articles have more than 1,000 coauthors. The article entitled ``International prevalence, recognition, and treatment of cardiovascular risk factors in outpatients with atherothrombosis” by Deepak L. Bhatt et al., and REACH Registry Investigators (2006) has 5,576 authors and was published in the \textit{Journal of the American Medical Association}. 

Given millions of articles, almost a billion references, and 254 subjects, any comparison and interpretation is challenging. Due to space limits, we aggregated the closely related subjects to a higher level scientific cluster and present the comparison at the higher cluster level. Note that all computations are still based on the article level. We only present the statistics and visualization at the cluster level, which such as median or mean are computed based on subjects of the same level. We identified 14 clusters based on the intensity of cross-citations and our knowledge of scientific disciplines. Specifically, we computed cross citation intensity between any two subjects by summing up the cross-citations between the two subjects and standardizing these values by the number of all articles in the two subjects. Given the intensity matrix, we adopted the graphical clustering approach \citep{wu2010gap} to form clusters according to the proximity measures, where subjects with high cross-citation intensity are grouped together, and subjects with low cross-citation intensity are separated using the elliptical separation algorithm and the property of a converging sequence of iteratively formed correlation matrices, \cite{chen2002generalized}. Next, we manually fine tuned the clustering by merging subjects with similar topics. We obtained 14 scientific clusters and presented a comparison among disciplines at the cluster level. Appendix \ref{app2} lists the subjects contained in each cluster as well as the corresponding number of articles, references and citations.

\begin{figure}[ht!]
		\begin{subfigure}[b]{0.5\textwidth}
		\centering
		\includegraphics[angle=90,page=4, trim=3cm 4cm 2.5cm 7cm, clip, width=1.1\textwidth, height=0.4\textheight ]{images/34.pdf}
		\includegraphics[trim=5.5cm 0cm 0cm 17cm, clip, width=1.1\textwidth, height=0.04\textheight]{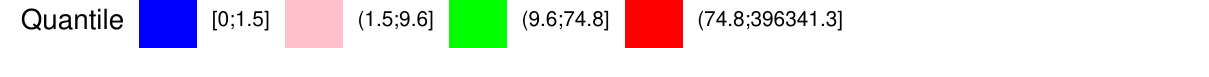}
		\caption{}
		\label{fig:heatmap}
	    \end{subfigure} 
		\begin{subfigure}[b]{0.5\textwidth}
		\centering
		\includegraphics[trim=22cm 0cm 0cm 0cm, clip,width=1.6\textwidth, height=0.45\textheight]{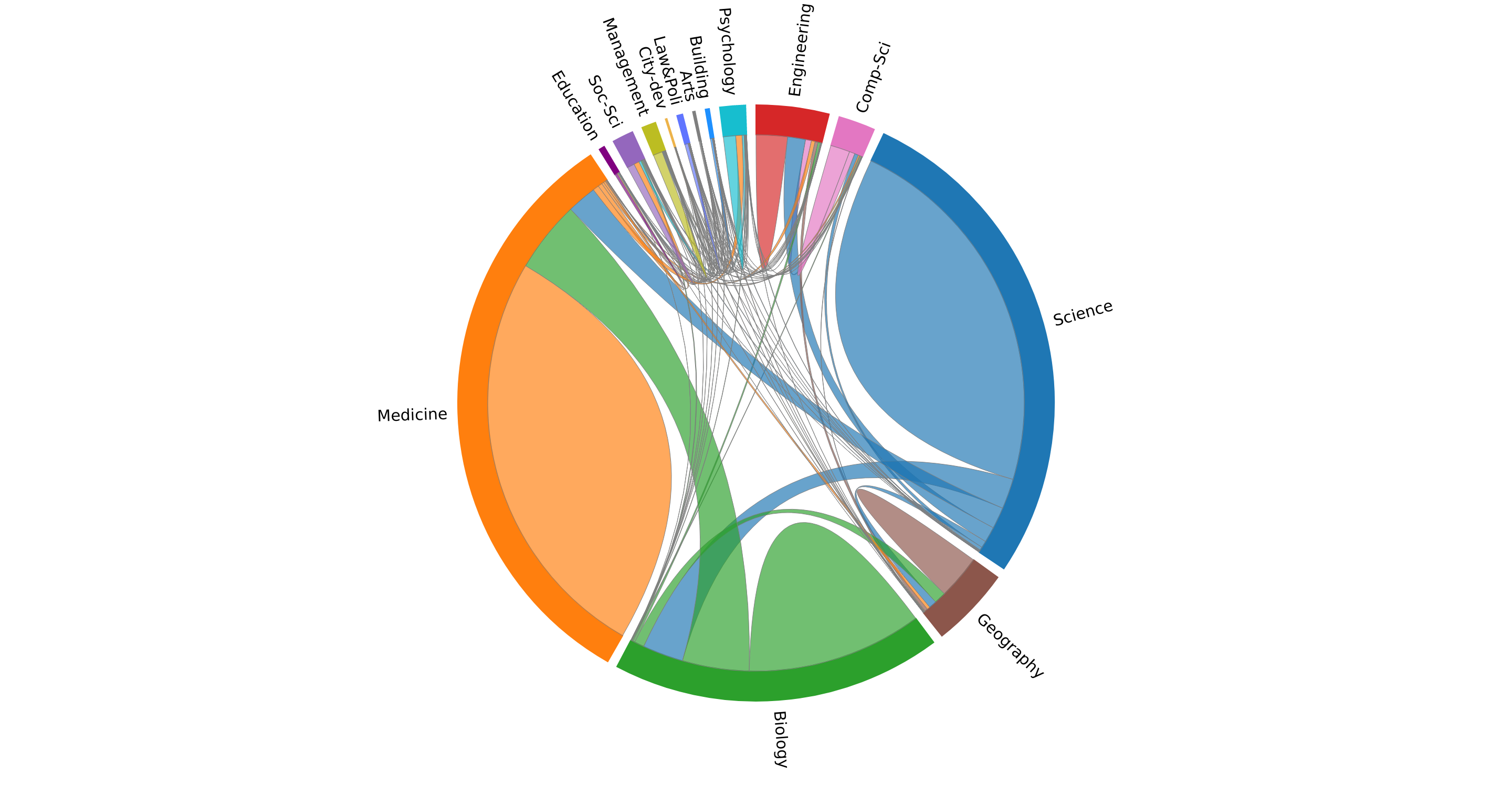}
		\caption{}
		\label{fig:chord}
	\end{subfigure}
	\caption{Panel (a) Heatmap of the intensity of cross subject citation. The quantiles show the intensity of the cross citations between subjects. Panel (b) Chord diagram for the cross-citations among the 14 clusters}
	\label{fig:heatmap_chord}
	\end{figure}

Figure \ref{fig:heatmap_chord} displays the cross-citation intensity matrix among the 254 subjects (panel \subref{fig:heatmap}) and the cross-citation chord diagram among the 14 clusters (panel \subref{fig:chord}). A visualization clustering package GAP\footnote{The software is downloaded from: \url{http://gap.stat.sinica.edu.tw/GAP/index.htm}} is used to arrange the 254 disciplines according to the proximity measures. In the intensity matrix, some cells along the diagonal demonstrate a high intensity of cross citations (coloured in red), with value $>$ 74.8, the upper quartile of intensity, whereas some cells representing e.g. Arts and Science off the diagonal have low a intensity of cross citations (coloured in blue), with value $<$ 1.5, the lower quartile of intensity. This is consistent with the
chord diagram in panel (\subref{fig:chord}), which also gives a good impression of the proportion of the clusters regarding the number of articles. About half of the articles in WoS belong to either the Medicine or Biology cluster. There is high cross-citation among subjects in the common cluster, where links are circled back to the same cluster, with the line thickness reflecting the strength of interdisciplinary cross citations. The chord diagram also shows that certain clusters such as Science and Biology do influence other clusters. Specifically, Science (blue area) is cited intensively by Biology, Medicine, Engineering, Computer Science and others, with blue linked projects to these areas. Biology (green area) is cited by Medicine and Geography. Medicine (orange area) is cited by Science and Biology. Other clusters, in contrast, have less cross-disciplinary citations.

\section{Method and Measure of ASP}
\label{sec:method}
\subsection{Eigenvector centrality}
We propose ASP to evaluate the scientific prestige of an article, based on eigenvector centrality or Pagerank. The idea of using eigenvector centrality to analyze a citation network is not new (see, e.g., \citealt{ma2008bringing}).
However, we are now able to compute Pagerank on the millions of node-graphs spanning all science disciplines in reasonable computational time. Articles together with the references build a mostly acyclic graph, where the direction of arcs is important. The challenge however exists in the boundary, where the leaves (newest publications) have no incoming arcs. To remedy this border effect, we performed our computations on the full graph from 1981 to 2020, but only looked at the results of articles from 1990 to 2015.

The \textit{ASP} of an article $i$ in a citation network of
size $N$ is defined as follows:
\begin{equation}\label{eq:ASP}
\text{\textit{ASP}}_i=(1-d)+d \sum_{j=1}^N \text{\textit{ASP}}_j L_{ij}/m_j
\end{equation}
where $L_{ij}=1$ if an article $j$ cites an article $i$ and $L_{ij}=0$ otherwise, $m_j=\sum_{k}L_{kj}$ is the total number of articles that $j$ links to. In other words, the ratio $L_{ij}/m_j$ denotes the fraction of references article $j$ has cited. The damping factor $d$ influences how much ``prestige'' of an article is passed on to the references. If $d$ is larger, more is passed on to the referenced (older) articles. As $d$ gets smaller, the benefit of being cited decreases. 
The minimum value of \textit{ASP} is $1-d$, which means that the article is not cited. We argue that an article that is never cited, or equivalently has the minimum value of \textit{ASP}, has negligible scientific impact. 

Present in matrix form, \textit{ASP} is eventually an eigenvector of a Markov matrix. Let
$$\text{\textit{ASP}}_{N\times 1}=
\begin{bmatrix}
    \text{\textit{ASP}}_1     \\
    \text{\textit{ASP}}_2     \\
    \vdots \\
    \text{\textit{ASP}}_N  
\end{bmatrix}, \ 
\text{\textbf{L}}_{N\times N}=
\begin{bmatrix}
    L_{11} & L_{12} & \dots & L_{1N}    \\
    L_{21} & L_{22} & \dots & L_{2N}    \\
    \vdots &\vdots    &   \ddots     & \vdots         \\
    L_{N1}  & L_{N2} & \dots & L_{NN}
\end{bmatrix}, \
\text{\textbf{M}}_{N\times N}=
\begin{bmatrix}
    m_{1} & 0 & \dots & 0    \\
    0 & m_{2} & \dots & 0    \\
    \vdots &\vdots    &   \ddots     & \vdots         \\
   0  & 0 & \dots & m_N
\end{bmatrix},
$$
and $\mathbf{\Omega}=\frac{1-d}{N}\mathbf{E}+d\mathbf{LM}^{-1}$ is a strongly connected Markov chain with a transition matrix $\mathbf{\Omega}^\top$, $\mathbf{E}$ is a $N\times N$ matrix of $1'$s and the damping factor $0<d<1$. From (\ref{eq:ASP}), we have
$$ \text{\textit{ASP}}=\mathbf{\Omega}  \text{\textit{ASP}},$$
where \textit{ASP} is the eigenvector of the matrix $\mathbf{\Omega}$ with an eigenvalue $1$. The matrix $\mathbf{\Omega}$ follows the Markov Chain with  
$$P \text{(go from j to i})=\begin{cases} (1-d)/N+d/m_j, & \mbox{if } \mbox{ j cites i} \\ (1-d)/N, & \mbox{if } \mbox{ j does not cite i} \end{cases},$$
which means that the chain moves from state j to state i with probability $(1-d)/N+d/m_j,$ if the paper j cites the paper i, and with probability $(1-d)/N,$ otherwise. The transition probability is the mixture of either randomly starting from a new article with the probability $1/N$, or follow one of the references of the paper j with the probability $1/m_j,$ respectively. If an article $j$ has many references, the probability of going from $j$ to a certain article $i$ in the reference becomes low.  

\subsection{Algorithm}
To solve the above equation system, we implement parallel Jacobi iterations on sparse data-structures, resulting in a steady-state solution for the ASP computation, see \cite{GolubVanLoan2013} and \cite{srivastava2019aitken}. The procedure is formulated in Algorithm \ref{Jacobi}. It begins by assigning an identical amount of prestige to each article. Next, this weight is redistributed in an iterative process whereby the articles transfer their attained weight to each other through the citations. The process ends when the difference between articles’ prestige values in consecutive iterations does not surpass a pre-established threshold. After setting up the data structures, the algorithm typically takes less than 2 seconds per iteration and converges in less than 20 iterations when running on an 8 core Intel Core i7-9700K CPU at 3.60GHz.

\begin{algorithm}[ht!]
    \caption{ASP computation}
    \hspace*{\algorithmicindent} \textbf{Input:} $d=0.5$, $N,$ $\mathbf{\Omega} \in \mathbb{R}^{N\times N},$ $\mathbf{E} \in \mathbb{R}^{N\times N},$  $\mathbf{M}\in \mathbb{R}^{N},$\\  \hspace*{\algorithmicindent}\textit{ASP}$^{(0)}\in \mathbb{R}^{N}$ initialized equal to 1, $\epsilon=0.01,$ $k=0$ \\
    \hspace*{\algorithmicindent} \textbf{Output:}  \textit{ASP}
    \begin{algorithmic}[1]
    \Procedure{Jacobi-Iteration:}{}\label{Jacobi}

    \State $\textbf{while} \ \max|\epsilon|\geq 0.01$
 
    \State \textit{ASP}$^{(k+1)} \gets (1-d)+d\mathbf{AM}^{-1}\times$\textit{ASP}$^{(k)}$
    \State $\mathbf{\epsilon}=$\textit{ASP}$^{(k+1)}-$\textit{ASP}$^{(k)}$ 
    \EndProcedure
    \end{algorithmic}
    \end{algorithm}

\subsection{Damping factor and citing window}\label{Sec:damping_factor}
There are two hyperparameters to choose in our algorithm. The damping factor $d$ decides how much of the incoming weight to a node is passed along to the referenced nodes. 
If a too-high damping factor is chosen, the oldest articles would receive most of the ASP, since they have no outgoing references within the data. When the damping factor is 1, the Markov chain matrix $\mathbf{\Omega}$ becomes irreducible, meaning article $j$ cannot reach article $i$ in a finite number of steps. For example, the articles published in later years cannot be reached (cited) by earlier published articles in the network. Therefore, if $d=1$, ASP converges to zero. If a too-low value  is chosen, all the weights would stay with the article, and very little would be conferred to the references. Moreover, as mentioned in the introduction, given the study period lasts 25 years from 1990 to 2015, it is reasonable to have a fixed citing window for a fair comparison for articles with different life lengths. 

The obvious question is what is a good damping factor, together with which citing window? By assuming that no subject should be better than another in terms of scientific contribution, we chose the hyperparameters that lead to the minimum variations among the 254 subjects. Specifically, we computed the average value of ASP in each subject. The difference is measured between the subject average ASP and the average ASP among all articles. We conduct the above computations for each year to avoid time impact. It shows that the choice of $d=0.5$ and a citing window of 5 years led to the minimum deviation among the scientific disciplines. This choice is also consistent with the fact that an article usually traces up to two consecutive articles \citep{chen2007finding}. In the following, we conduct the citation analysis based on that choice in our study. 
Appendix \ref{app3} details the choice. 

\section{ASP}
\label{sec:asp}
\begin{figure}[ht!]
	\begin{subfigure}[b]{0.5\linewidth}
		\centering
		\includegraphics[width=1.1\textwidth, height=0.4\textheight]{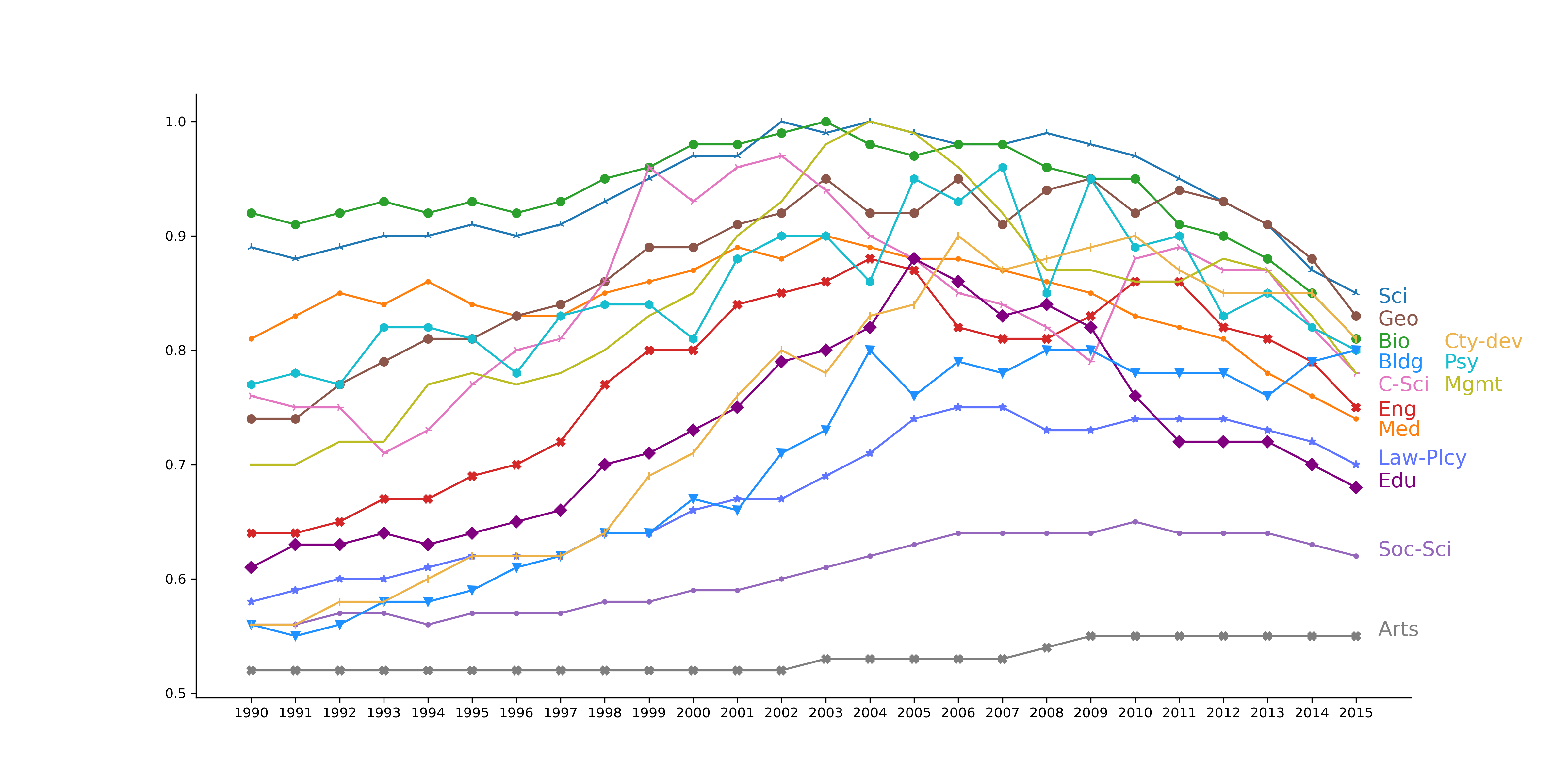}
		\caption{ASP}
		\label{fig:ASP_avg}
	\end{subfigure}
	\begin{subfigure}[b]{0.5\linewidth}
	\centering
	\includegraphics[width=1.1\textwidth, height=0.4\textheight]{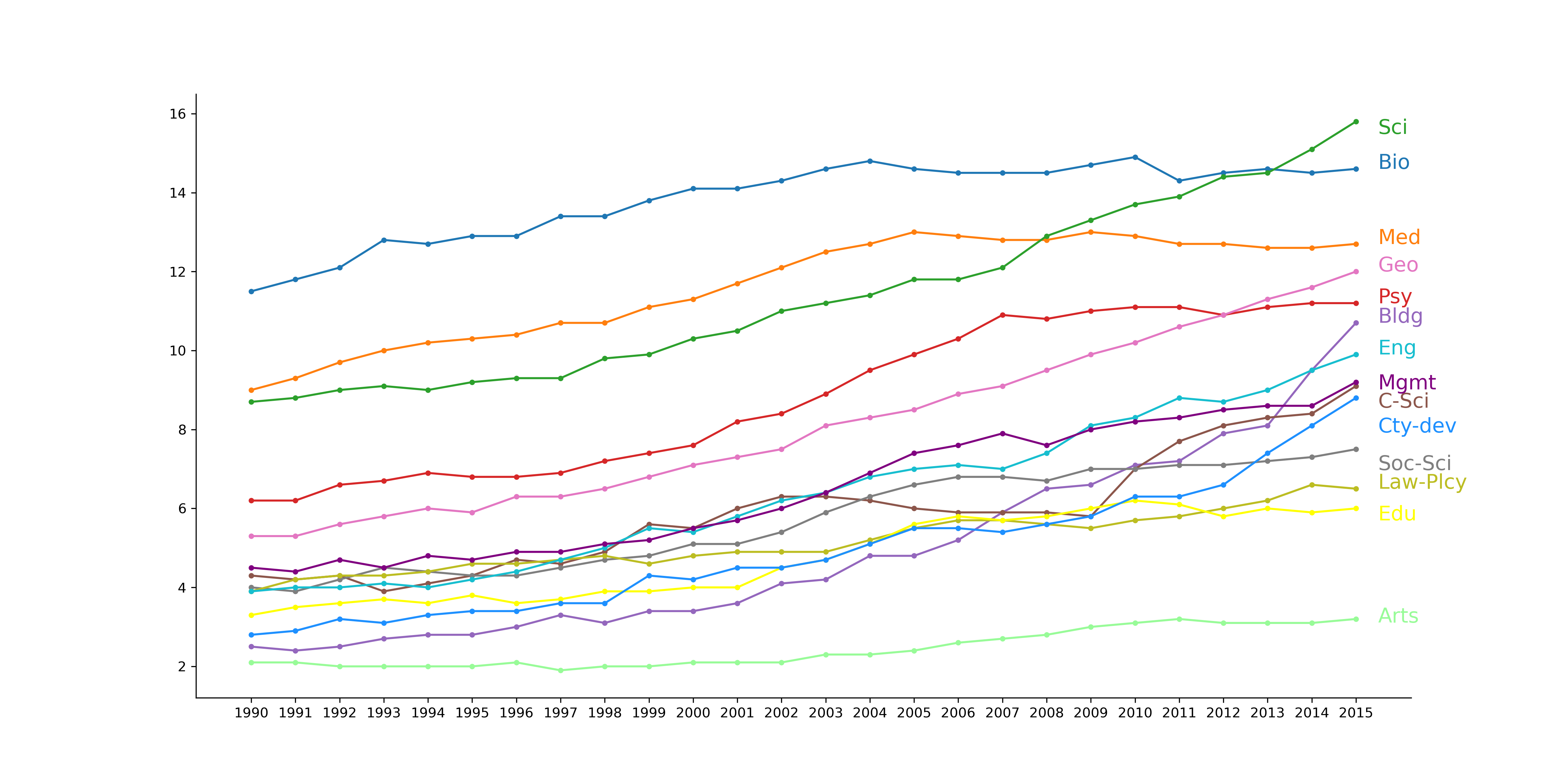}
	\caption{\citcount}
	\label{fig:ncit_avg}
\end{subfigure}
	\caption{Average values of ASP and \citcount for 14 clusters over years between 1990 and 2015}
\label{fig:average_ASP}
\end{figure}

We summarize the statistics and distributional properties of ASP in this section. Table \ref{tab:my_label_cluster_distr} lists some statistics of ASP and \citcount summarized for the 14 clusters. Without exception, both metrics are right-skewed distributed and have extremely large values. For several clusters, the mean is larger than the upper quartile (Q3), indicating the existence of extreme large values. In terms of ASP, Biology, Science, Medicine, and Psychology lead with the largest average values and there is a minor difference up to the upper quartile (Q3) among the top 4 clusters. 
In contrast, Arts and Social Science have the lowest ASP, with 50-75\% articles never being cited. Figure \ref{fig:average_ASP} panel (\subref{fig:ASP_avg}) presents the dynamic evolution of the average ASP of the 14 clusters from 1990 to 2015\footnote{The median ASP for 14 clusters between 1990 and 2015 are piecewise constant over years and overlapping between clusters, see Figure \ref{fig:med_asp} in Appendix \ref{app2}.}.
Biology and Science are superior performers over time, with stable and large values. Medicine, though with the largest number of papers and citations, has been overtaken by Geography, City Development, Computer Science, Management, and Engineering in recent years. Meanwhile, Law \& Policy, Social Science, and Arts attract attention with  steadily improved performance. Nevertheless, their overall impacts are still marginal compared to others. Education and Management are special, with a hump around 2005-2009, which may represent their golden age due to the education reforms on pedagogy such as the ``No Child Left Behind Act'' introduced by the U.S. Act of Congress in the beginning of 2000 or the development E-commerce including online marketing and sales in management. Except for a few of these variations, there is generally a stable relative performance among the different disciplines over time.

The rank of clusters differs in terms of \citcount. When only direct citations are considered, Psychology and Medicine upgrade to \#2 and \#3, see Table  \ref{tab:my_label_cluster_distr} and Figure \ref{fig:average_ASP} panel (\subref{fig:ncit_avg}). There are also larger deviations in magnitude among the clusters. The difference between \citcount and ASP can be further illustrated using a single article as an example. The article “A short history of SHELX” by George Michael Sheldrick (2008), which was published in \textit{Acta Crystallographica Section A: Foundations of Crystallography} has the maximum
ASP of 1,570.09. The article belongs to the Science cluster, with the subjects of Chemistry, Multidisciplinary, and Crystallography. Although it received 803 \citcount only, which is much less than the highest \citcount, 58,002, its scientific prestige is higher due to indirect citations. Specifically, this article was cited by other articles with high impact, which eventually enhanced its influence in the citation network. For example, it was cited by the article ``Structure validation in chemical crystallography'' by Anthony L. Spek (2009) with the ASP of 234.37 and \citcount=10,412, and article ``OLEX2: a complete structure solution, refinement and analysis program'' by Oleg V. Dolomanov et al. (2009) with the ASP of 112.43 and \citcount=10,378. When counting all articles directly or indirectly linked to the article, the citation counts is more than 80,000 . In other words, while a direct count of citations for the article is merely 803, there is an impact via indirect citations too. And the impact of the indirect citation is only considered in the computation of ASP. 

\begin{table}[ht!]
    \centering
\scalebox{0.8}{\begin{tabular}{l|rrrrrr|rrrrrr}
\toprule
{} & \multicolumn{6}{c}{\citcount} & \multicolumn{6}{|c}{ASP} \\
\midrule
Cluster &    Min &  Q1 &   Median &   Mean &   Q3 &      Max &      Min &   Q1 &   Median &  Mean &   Q3 &      Max \\
\midrule
Science          &    0 &  2 &   9 &  25.64 &  24 &  53,341 &      0.5 &  0.53 &  0.63 &  0.92 &  0.87 &  1,570.09 \\
Medicine         &    0 &  3 &  11 &  27.73 &  29 &  28,134 &      0.5 &  0.54 &  0.64 &  0.91 &  0.89 &   550.59 \\
Biology          &    0 &  6 &  16 &  36.09 &  38 &  58,002 &      0.5 &  0.56 &  0.67 &  0.94 &  0.93 &  1,097.45 \\
Engineering      &    0 &  0 &   2 &  12.18 &  10 &  13,912 &      0.5 &  0.50 &  0.55 &  0.84 &  0.76 &   507.67 \\
Social Science   &    0 &  0 &   0 &   5.78 &   2 &  10,326 &      0.5 &  0.50 &  0.50 &  0.62 &  0.55 &   168.35 \\
Geography        &    0 &  3 &  11 &  24.50 &  28 &  19,894 &      0.5 &  0.54 &  0.64 &  0.86 &  0.88 &   362.37 \\
Computer Science &    0 &  0 &   2 &  11.40 &   8 &  33,120 &      0.5 &  0.50 &  0.55 &  0.88 &  0.73 &   768.86 \\
Arts             &    0 &  0 &   0 &   0.98 &   0 &   1,228 &      0.5 &  0.50 &  0.50 &  0.55 &  0.50 &    88.54 \\
Management       &    0 &  0 &   4 &  20.59 &  16 &  13,700 &      0.5 &  0.50 &  0.56 &  0.90 &  0.79 &   280.47 \\
Psychology       &    0 &  1 &   8 &  27.85 &  27 &  23,345 &      0.5 &  0.51 &  0.61 &  0.91 &  0.85 &   372.76 \\
Law and Policy   &    0 &  0 &   1 &   8.42 &   7 &   2,410 &      0.5 &  0.50 &  0.51 &  0.72 &  0.68 &   130.34 \\
Education        &    0 &  0 &   2 &   9.45 &   9 &   6,063 &      0.5 &  0.50 &  0.54 &  0.76 &  0.74 &   105.90 \\
Building         &    0 &  0 &   1 &  11.53 &  11 &   2,593 &      0.5 &  0.50 &  0.54 &  0.82 &  0.79 &    68.41 \\
City Development &    0 &  0 &   3 &  12.19 &  12 &   1,457 &      0.5 &  0.50 &  0.57 &  0.81 &  0.79 &    50.59 \\
\midrule
Total  & 0 &	1 & 	7 &	22.82& 22 & 58,002 & 0.5 &	0.51 &	0.60	& 0.87 & 0.83 &	1570.09 \\ 

\bottomrule
\end{tabular}}
    \caption{Distribution of \citcount and ASP at cluster level}
    \label{tab:my_label_cluster_distr}
\end{table}

\begin{figure}[ht!]
	\begin{subfigure}[b]{0.5\linewidth}
		\centering
		\includegraphics[width=1.1\textwidth, height=0.4\textheight]{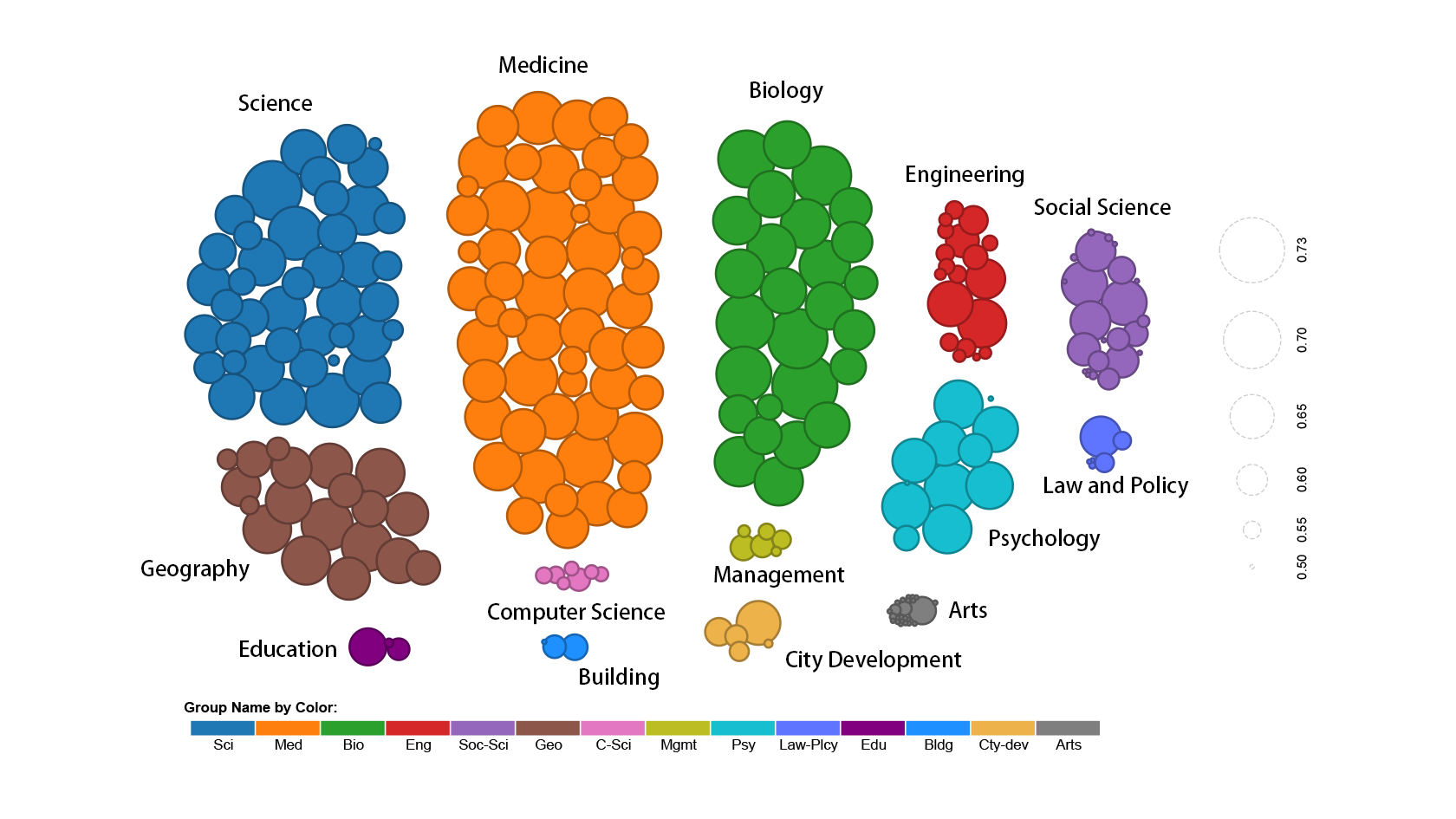}
		\caption{ASP}
		\label{fig:ASP_bubble}
	\end{subfigure}
	\begin{subfigure}[b]{0.5\linewidth}
	\centering
	\includegraphics[width=1.1\textwidth, height=0.4\textheight]{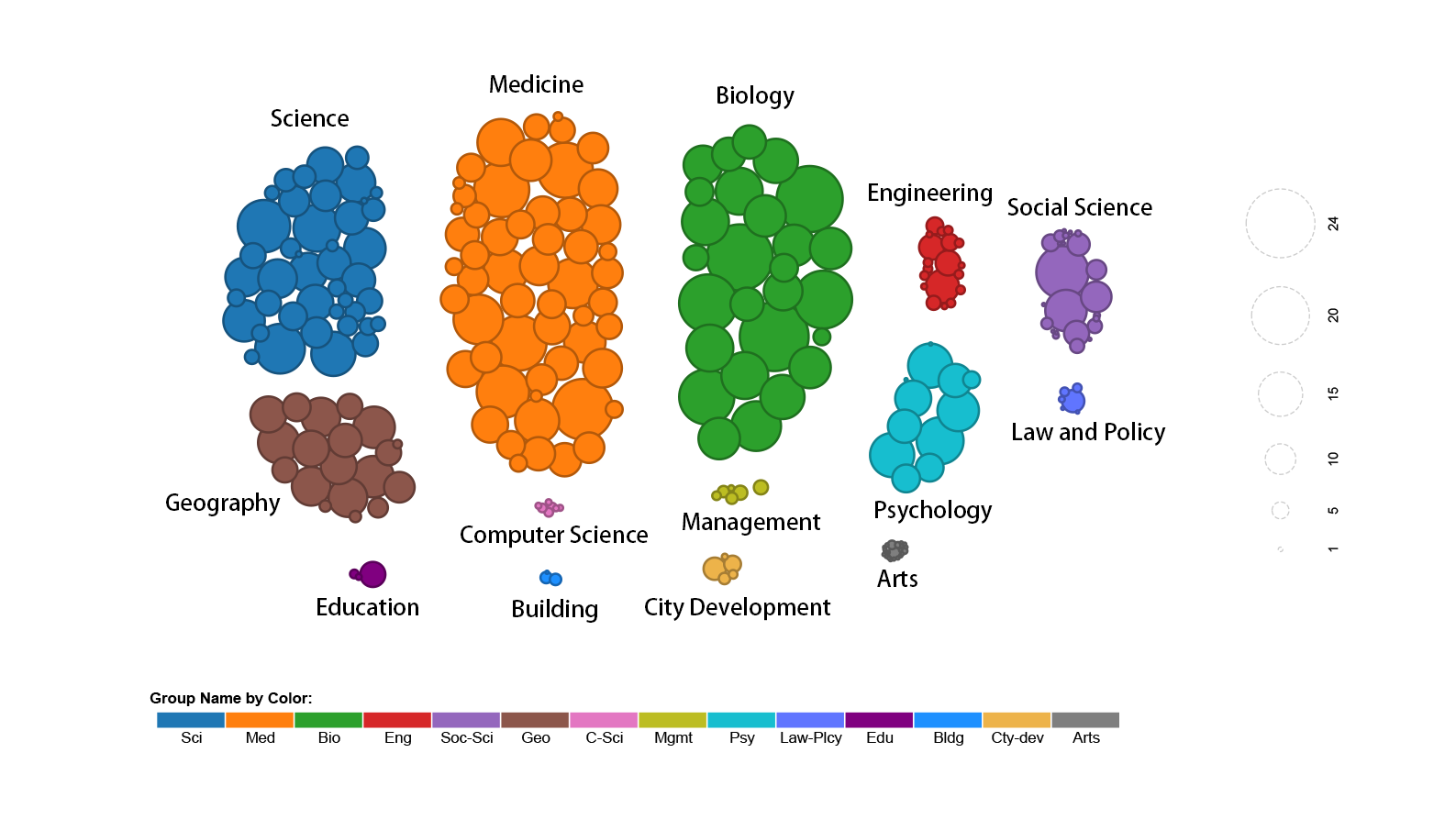}
	\caption{\citcount}
	\label{fig:ncit_bubble}
\end{subfigure}
	\caption{Medians of ASP and \citcount for all the 254 subjects}
	\label{tab:ASP_bubble}
\end{figure}

To further illustrate the difference between the two citation metrics at article level, Figure \ref{tab:ASP_bubble} presents the medians of ASP and \citcount for all 254 subjects, grouped into the 14 clusters. It shows that Biology, Science, Medicine, Geography, and Psychology are more influential with larger ASP. Arts and Management are usually less influential with smaller ASP. Note that Computer Science has a surprisingly small ASP. This is due to the data restrictions on articles in our study as conference proceedings is the main publication stream in Computer Science. There is a larger variation in Engineering, Social Science, and City Development, with both very large and very small medians. At first glance, \citcount displays analogous distribution among the 254 subjects. It is, however, interesting to note that the relative relation of subjects differs in terms of \citcount. The difference from cluster to cluster, for example Science vs. Arts, becomes dominant, compared to that in ASP. Moreover, we observe much bigger variations within clusters. For example, the subjects within Science have a larger difference when \citcount are used as metric. This implies that ASP is a more stable metric for different disciplines.

\begin{figure}[ht!]
	\begin{subfigure}[b]{0.5\linewidth}
		\centering
		\includegraphics[width=1.1\textwidth, height=0.4\textheight]{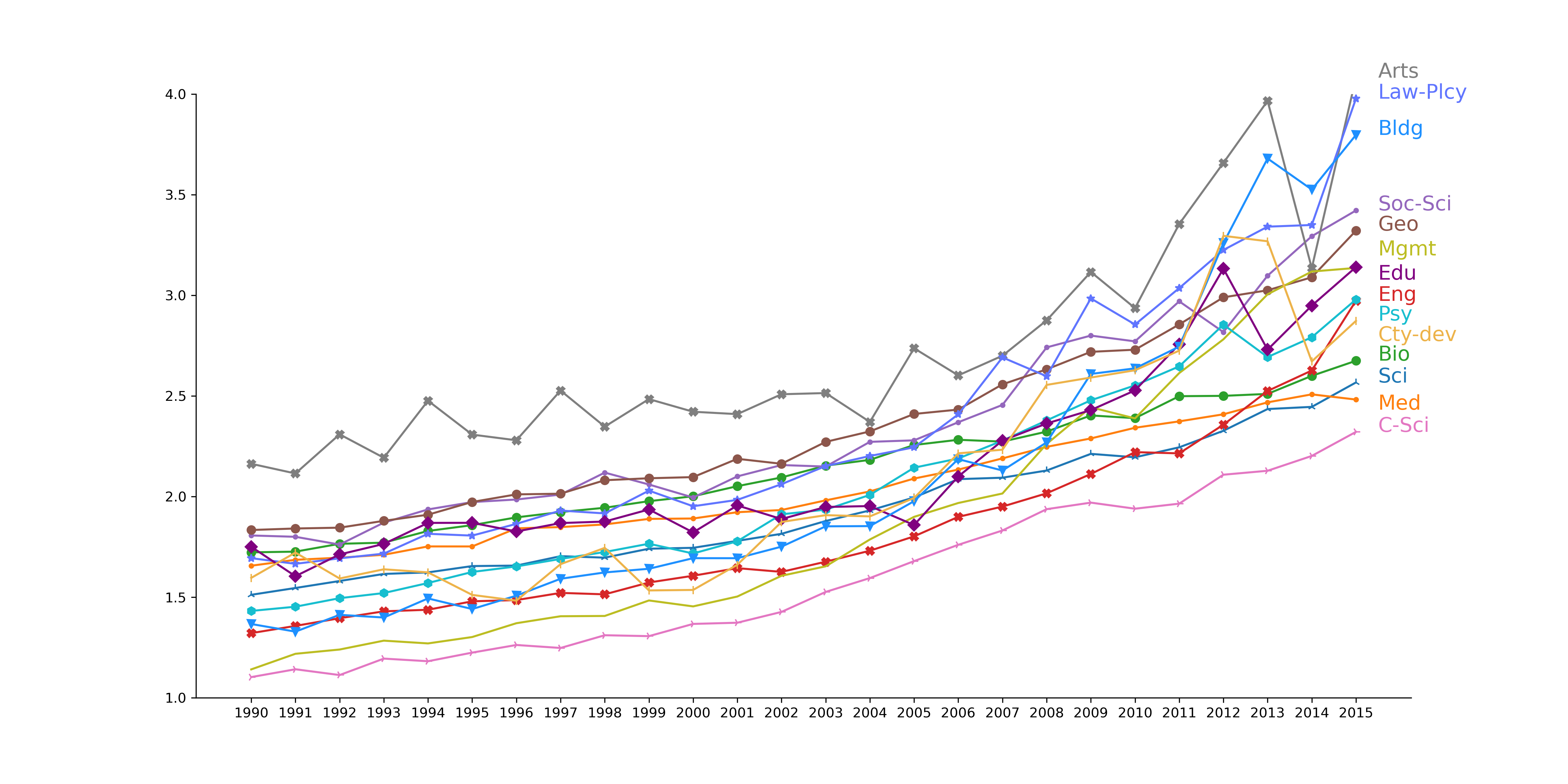}
		\caption{ASP}
		\label{fig:tiasp}
	\end{subfigure}
	\begin{subfigure}[b]{0.5\linewidth}
	\centering
	\includegraphics[width=1.1\textwidth, height=0.4\textheight]{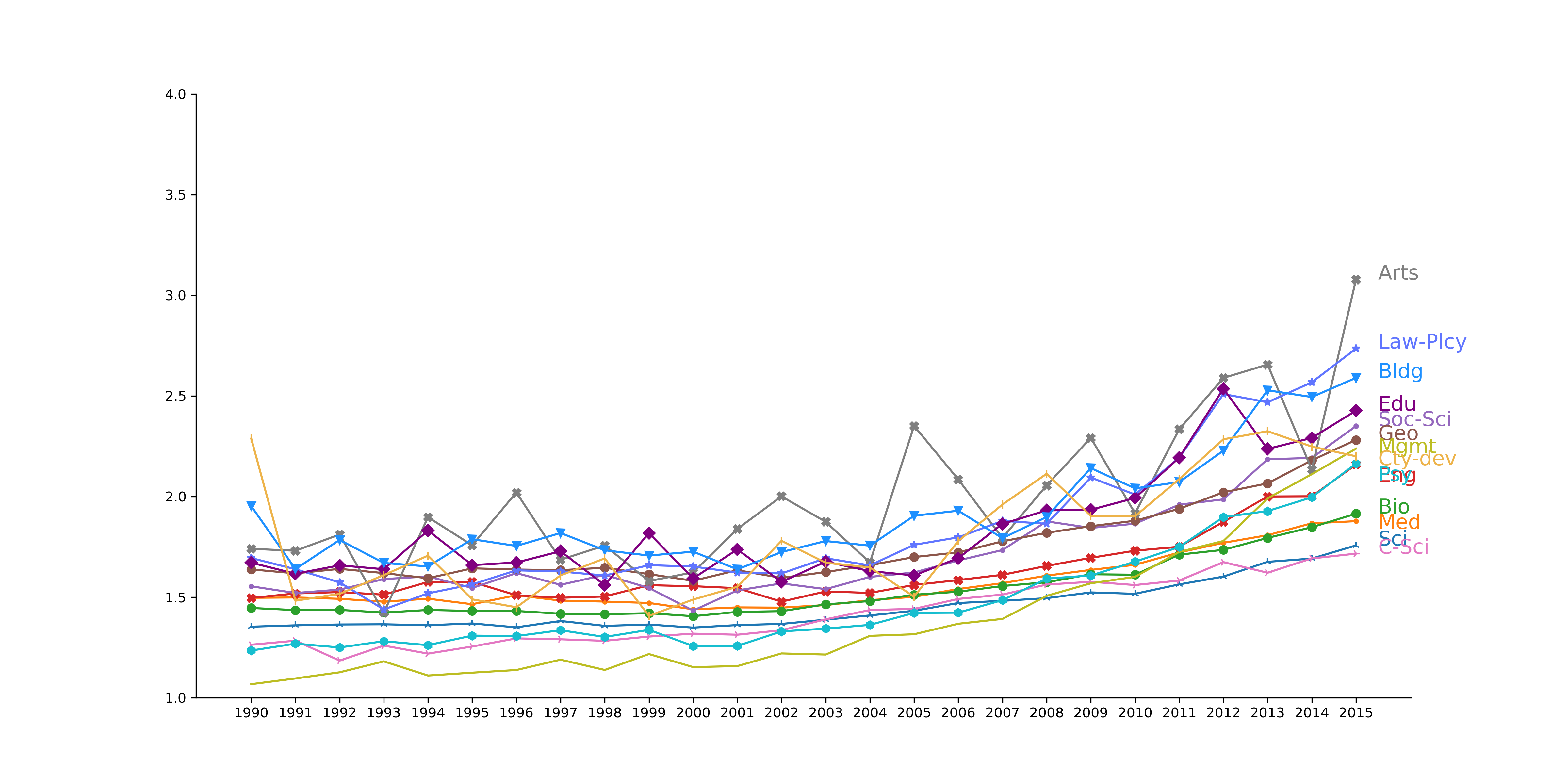}
	\caption{\citcount}
	\label{fig:tinc}
\end{subfigure}
	\caption{Tail index for the ASP and \citcount published between 1990 and 2015}
	\label{fig:TI}
\end{figure}

Analogous to the Pareto principle in economics, right skewed distribution implies that, in the citation network, very few articles have the most scientific influence or citations, while the rest are rarely cited or not cited at all. We use the Pareto distribution to approximate the tail behaviour of ASP, whose probability density function is defined as: 
$$p(x)= \alpha x_{\min}^\alpha/x^{1+\alpha}, \ x \geq x_{\min}>0.$$
The shape parameter $\alpha$, also known as the tail index, describes the heaviness of the tail. The larger the tail index, the smaller the proportion of extreme values and vice versa for thinner tails. We estimate the tail index with a threshold of $x_{\min}$ to be the top 10\% percentile of ASP, which is 1.88. Figure \ref{fig:TI} presents the series of tail indexes estimated based on ASP (panel \subref{fig:tiasp}) and \citcount (panel \subref{fig:tinc}) for the 14 clusters from 1990 to 2015. We found that there is an increasing trend in ASP, indicating the increasing influence of a few top papers dominating the scientific contribution in the science. Given the total number of publications is increasing, it also means that the ratio of irrelevant publications is rising. Arts, City Development, and Law \& Policy are less extreme; one would expect increased imbalance severity with relatively fewer outstanding articles in the cluster. In contrast, long tails are common in Engineering, Science, and Computer Science. In contrast, both the magnitude and the increase in trend are weaker when \citcount are considered. While the average tail index of ASP increases from 1.58 to 2.63 over the years, it varies from 1.43 to 1.88 for \citcount, and most clusters remain below 2, except Arts after 2005 and City Development in 1990. This implies that, as a metric, ASP leads to less extreme distributions than \citcount.

\section{ASP and the alternative metrics}
\label{sec:asp_ncit}
\subsection{ASP and \citcount}
\citcount is possibly the most commonly used evaluation metric for individual articles. Though direct and convenient, citation count has been criticized for several shortcomings. As mentioned, the citation metric is not comparable across disciplines where the frequency of citation differs. For example, physics articles are published at a much higher frequency and are more likely to have higher citation counts than mathematics articles. Also, self-citation or community-citation can easily abuse citation counts. Even excluding self-citations, it is sometimes still unclear whether an article is more important judging with higher citation counts alone. First, it depends on the type of article. A survey article usually includes a lot references on average and will possibly be cited more often. In some sense, a survey article may dilute the importance (if reflected by citation counts) of some articles, as the latter would be cited as a whole as in the survey article. Second, it also depends on the wide recognition of the work. Newton's gravitational law would be directly used without citing the original work. A recent example is Roger Penrose’s work “Gravitational Collapse and Space-Time Singularities” published in 1965. This article later won Penrose the “2020 Nobel Prize in Physics”,  but received in total only 153 \citcount after publication up to December 2020 according to MathSci. More importantly, it lacks sufficient empirical evidence on how reliable the citation metric is at the article level.

Given that ASP is computed based on citation counts but also considers the sequential impact of the article via indirect citations, i.e., influence of other articles that cite, questions arise: 1) what is the relationship between the two citation metrics and 2) which metric is more reliable and under which situations.
We computed the Pearson correlations between ASP and \citcount. In the computation, we remove articles without citations, which correspond to about 22.53\% articles as the values of \citcount are always zero and ASP always 0.5, leading to meaningless perfect correlation. Appendix \ref{app4} presents the statistics of the articles without citation. 

\begin{figure}[ht!]
    \centering
    \begin{subfigure}[b]{0.5\textwidth}
        \centering
        \includegraphics[width=1\linewidth, height=0.35\textheight]{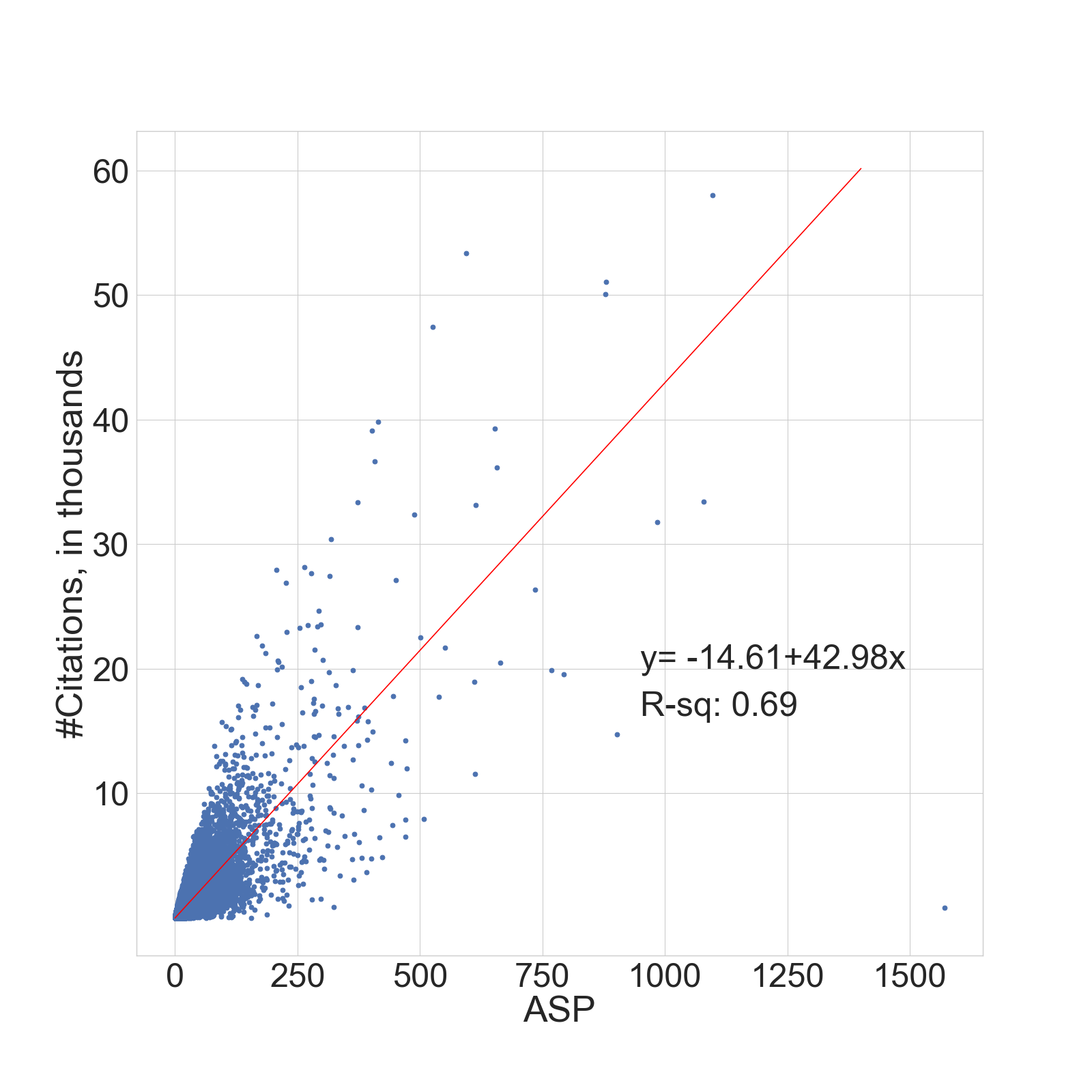}
        \caption{}
        \label{fig:ASP_ncit_scatter}
    \end{subfigure}%
    \begin{subfigure}[b]{0.5\textwidth}
        \centering
        \includegraphics[width=1\linewidth, height=0.35\textheight]{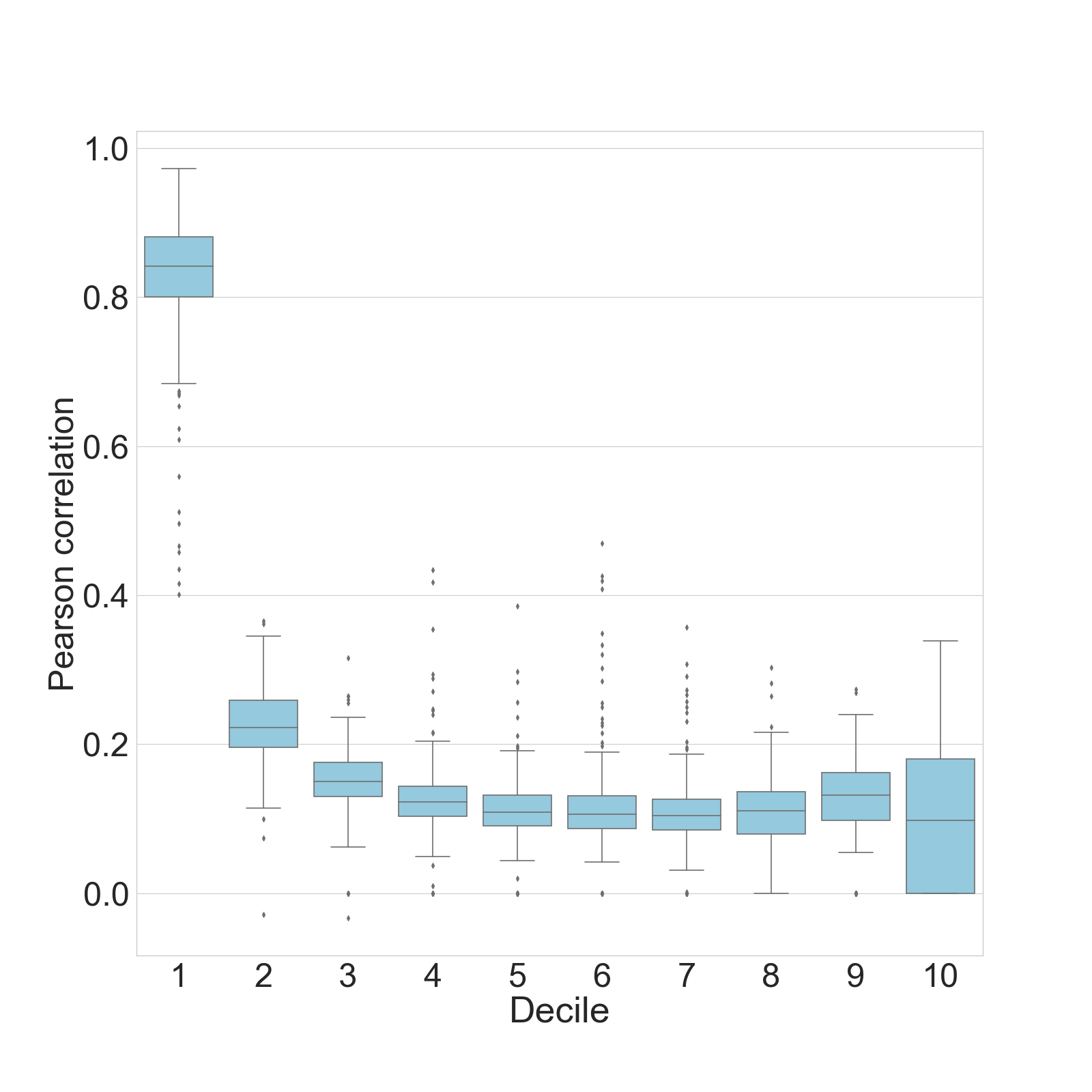}
        \caption{}
        \label{fig:ASP_ncit_Pearson10}
       
    \end{subfigure}
\caption{Panel (a). Scatterplot of \citcount vs the ASP and the fitted line. Panel (b). Pearson correlation of ASP and \citcount in 10 groups and over years between 1990 and 2015. Each group includes the correlation values with respect to subjects and deciles over years. Each decile is obtained by dividing articles in each subject into 10 equal groups according to their sorted \citcount}
 \label{fig:Scatter_corr}
 \end{figure}

We found that although the articles with the top 10\% \citcount, i.e., the articles with high citations, have similar ranks with ASP, the remaining 90\% articles differ significantly. It means it would be relatively safe to evaluate the scientific prestige of articles with either ASP or \citcount, but only for the top 10\% articles, i.e., on average with more than 24 
citations\footnote{The 90\% of all the articles have less than 11 citations. For articles with at least 1 citation, the 90\% quantile is 24}. However, it would be tricky by mixing the two metrics for the remaining articles.
Figure \ref{fig:Scatter_corr} panel (\subref{fig:ASP_ncit_scatter}) presents the scatterplot of \citcount vs ASP. First glance shows strong connection with a correlation coefficients of 0.82. One would expect that there is little difference of the two metrics for evaluating an article's prestige. Given the long tails of both metrics and the sensitivity of the correlation coefficients to outliers, we divided articles into deciles according to their sorted \citcount in each cluster, after removing the non-cited articles. The first group contains the top 10\% articles with the highest \citcount in each cluster, and the last group (\#10) has the last 10\% of articles with the lowest \citcount for each cluster. 
The boxplot of the Pearson correlation coefficients between ASP and \citcount is displayed in panel (\subref{fig:ASP_ncit_Pearson10}) for each of the 10 groups. Except the top 10\% articles have a high correlation at 0.804, the remaining 90\% articles have, on average, correlations below 0.2. The correlation, in general, drops further when the decile increases. There is generally a similar pattern for different clusters. Appendix \ref{app5} presents the correlation coefficients of decile groups for each of the 14 clusters over time, where correlation is high among the top 10\% articles, and low for the rest.

\subsubsection{Coincidence among the ``top'' articles}
To verify the relationship between the two metrics, we select some individual articles to perform a detailed investigation. Table \ref{tab:my_label4} lists the top 20 articles in each metric and their corresponding ranks. There are 10 articles that appear in both top 20 rankings, including the article with the maximum \citcount of 58,002 (ASP of 1097.44 ranked \#2). Meanwhile, there are also articles, within the top 10\%, exhibit ``big'' differences in ranks. The article ``A short history of SHELX” with ASP=1,570.09 and \citcount=803 is ranked as 1st according to its ASP but $33,229^{\text{th}}$ for its \citcount, which is still in the top 10\%. In contrast, the article ``Hallmarks of Cancer: The Next Generation” with a relatively lower ASP=206.64 but larger \citcount=27,946 is ranked $195^{\text{th}}$ according to its ASP and $18^{\text{th}}$ according to its \citcount. The comparison reconfirms the strong correlation of the two metrics for highly influential and highly cited articles within the top group. 

\begin{table}[ht!]
	\centering
	\caption{The sets of article with either highest 20 ASP or highest 20 \citcount in WoS dataset, where $n$ is the rank ASP and $k$ is the rank \citcount.}
	\centering
	\resizebox{1\linewidth}{!}{
\begin{tabular}{rrrrlrll}
\toprule
  n &        ASP &      k &   \citcount &                                              Title &  Year &                                             Source &           Cluster \\
\midrule
  1 &  1,570.09 &  33,229 &    803 &                           A short history of SHELX &  2008 &                          Acta Crystallogr. Sect. A &           Science \\
  2 &  1,097.44 &      1 &  58,002 &                  BASIC LOCAL ALIGNMENT SEARCH TOOL &  1990 &                                      J. Mol. Biol. &           Biology \\
  3 &  1,079.32 &     11 &  33,421 &     Generalized gradient approximation made simple &  1996 &                                   Phys. Rev. Lett. &           Science \\
  4 &   983.73 &     15 &  31,777 &           HELICAL MICROTUBULES OF GRAPHITIC CARBON &  1991 &                                             Nature &           Science \\
  5 &   902.72 &     87 &  14,720 &  DENSITY-FUNCTIONAL THERMOCHEMISTRY .3. THE ROL... &  1993 &                                     J. Chem. Phys. &           Science \\
  6 &   879.75 &      3 &  51,060 &  Gapped BLAST and PSI-BLAST: a new generation o... &  1997 &                                 Nucleic Acids Res. &           Biology \\
  7 &   878.72 &      4 &  50,090 &  CLUSTAL-W - IMPROVING THE SENSITIVITY OF PROGR... &  1994 &                                 Nucleic Acids Res. &           Biology \\
  8 &   793.63 &     46 &  19,549 &  PHASE ANNEALING IN SHELX-90 - DIRECT METHODS F... &  1990 &                          Acta Crystallogr. Sect. A &           Science \\
  9 &   768.86 &     44 &  19,875 &                            SUPPORT-VECTOR NETWORKS &  1995 &                                       Mach. Learn. &  Computer Science \\
 10 &   735.08 &     23 &  26,342 &  Analysis of relative gene expression data usin... &  2001 &                                            Methods &           Biology \\
 11 &   663.75 &     40 &  20,492 &      Collective dynamics of `small-world' networks &  1998 &                                             Nature &           Science \\
 12 &   656.95 &     10 &  36,177 &  Processing of X-ray diffraction data collected... &  1997 &                                   Methods Enzymol. &           Biology \\
 13 &   652.29 &      7 &  39,269 &  Electric field effect in atomically thin carbo... &  2004 &                                            Science &           Science \\
 14 &   613.32 &     13 &  33,120 &                                     Random forests &  2001 &                                       Mach. Learn. &  Computer Science \\
 15 &   611.86 &    145 &  11,543 &  Histograms of oriented gradients for human det... &  2005 &                                     PROC CVPR IEEE &  Computer Science \\
 16 &   611.01 &     50 &  18,940 &  NavOptim coding: Supporting website navigation... &  2004 &  IEEE/WIC/ACM INTERNATIONAL CONFERENCE ON WEB I... &  Computer Science \\
 17 &   594.11 &      2 &  53,341 &  Efficient iterative schemes for ab initio tota... &  1996 &                                       Phys. Rev. B &           Science \\
 18 &   550.58 &     34 &  21,699 &  THE MOS 36-ITEM SHORT-FORM HEALTH SURVEY (SF-3... &  1992 &                                          Med. Care &          Medicine \\
 19 &   538.72 &     56 &  17,769 &            Emergence of scaling in random networks &  1999 &                                            Science &           Science \\
 20 &   525.73 &      5 &  47,428 &  CONTROLLING THE FALSE DISCOVERY RATE - A PRACT... &  1995 &            J. R. Stat. Soc. Ser. B-Stat. Methodol. &           Science \\
 \midrule
   35 &   413.90 &   6 &  39,833 &                    PROJECTOR AUGMENTED-WAVE METHOD &  1994 &                             Phys. Rev. B &           Science \\
 38 &   401.62 &   8 &  39,092 &  From ultrasoft pseudopotentials to the project... &  1999 &                             Phys. Rev. B &           Science \\
  36 &   407.43 &   9 &  36,665 &  Cutoff Criteria for Fit Indexes in Covariance ... &  1999 &                    Struct. Equ. Modeling &           Science \\
 52 &   372.21 &  12 &  33,375 &  Efficiency of ab-initio total energy calculati... &  1996 &                      Comput. Mater. Sci. &           Science \\
 23 &   488.79 &  14 &  32,395 &  The CLUSTAL\_X windows interface: flexible stra... &  1997 &                       Nucleic Acids Res. &           Biology \\
 73 &   318.24 &  16 &  30,379 &  MEGA6: Molecular Evolutionary Genetics Analysi... &  2013 &                         Mol. Biol. Evol. &           Biology \\
 119 &   264.45 &  17 &  28,134 &  Preferred reporting items for systematic revie... &  2009 &              BMJ-British Medical Journal &          Medicine \\
 195 &   206.64 &  18 &  27,946 &           Hallmarks of Cancer: The Next Generation &  2011 &                                     Cell &           Biology \\
 109 &   277.70 &  19 &  27,665 &  Fast and accurate short read alignment with Bu... &  2009 &                           Bioinformatics &           Biology \\
  75 &   315.84 &  20 &  27,461 &           Measuring inconsistency in meta-analyses &  2003 &                              Br. Med. J. &          Medicine \\
 
 \bottomrule
\end{tabular}	}
	\label{tab:my_label4}
\end{table}

As another example of the high correlation between the top articles, we consider the publication records of Nobel laureates in Physics, Chemistry, and Physiology or Medicine, in a total of 24 articles, from 1990 to 2015. The list is retrieved from the Harvard Nobel prize papers database by \cite{li2019dataset}. We found, among the Nobel prize winning papers, 21 articles are ranked in top 1\% in both \citcount and ASP and 3 are higher ranked according to ASP, in the $95^{\text{th}},$ $90^{\text{th}},$ and $65^{\text{th}}$ percentiles, and lower ranked according to \citcount, in the $72^{\text{th}},$ $85^{\text{th}},$ and $59^{\text{th}}$ percentiles, supporting the argument that both metrics are reliable for evaluating the prestige of articles, although ASP provides more accurate rankings than \citcount when the articles are not highly cited, see Table \ref{tab:my_label_nobel}. 

\begin{table}[ht!]
	\centering
	\caption{The 24 Nobel Prize winning papers, ranked by ASP with percentile (\%ile) in each metric.}

	\resizebox{1\linewidth}{!}{
	\begin{tabular}{rrrrlrll}
	\toprule
 ASP \%ile &       ASP &           \citcount \%ile &   \citcount &                                              Title &  Year &            Source &  Cluster  \\
	\midrule 
	99 &  652.29 &        99 &  39,269 &  Electric field effect in atomically thin carbo... &  2004 &           Science &  Science \\      99 &  158.86 &       99 &  11,622 &  Observational evidence from supernovae for an ... &  1998 &        Astron. J. &  Science \\
	 99 &  137.06 &       99 &  13,254 &  Induction of pluripotent stem cells from mouse... &  2006 &              Cell &  Biology  \\      99 &  134.14 &       99 &   8,981 &  Potent and specific genetic interference by do... &  1998 &            Nature &  Science  \\    99 &  111.28 &       99 &  10,492 &  Induction of pluripotent stem cells from adult... &  2007 &              Cell &  Biology  \\    99 &   81.66 &      99 &   3799 &  Evidence for oscillation of atmospheric neutrinos &  1998 &  Phys. Rev. Lett. &  Science  \\    99 &   71.81 &      99 &   4,157 &  Quantum phase transition from a superfluid to ... &  2002 &            Nature &  Science  \\   99 &   50.21 &      99 &   2,421 &  The dorsoventral regulatory gene cassette spat... &  1996 &              Cell &  Biology  \\   99 &   35.80 &      99 &   2,439 &  The complete atomic structure of the large rib... &  2000 &           Science &  Science \\ 99 &   31.67 &      99 &   1,800 &  Discovery of a supernova explosion at half the... &  1998 &            Nature &  Science  \\  99 &   30.91 &      99 &   2,023 &  Direct evidence for neutrino flavor transforma... &  2002 &  Phys. Rev. Lett. &  Science \\    99 &   29.63 &     99 &   1,301 &  Observing the progressive decoherence of the '... &  1996 &  Phys. Rev. Lett. &  Science  \\  99 & 25.90 &     99 &    840 &  Direct link between microwave and optical freq... &  2000 &  Phys. Rev. Lett. &  Science  \\  99 &   22.61 &     99 &    922 &  Generation of nonclassical motional states of ... &  1996 &  Phys. Rev. Lett. &  Science \\  99 &   21.93 &     99 &   1,480 &  The structural basis of ribosome activity in p... &  2000 &           Science &  Science  \\  99  &   18.11 &     99 &   1,607 &  Visualizing secretion and synaptic transmissio... &  1998 &            Nature &  Science  \\  99 &   17.28 &      99 &   1,676 &  Microstructure of a spatial map in the entorhi... &  2005 &            Nature &  Science \\  99 &   16.99 &     99 &    574 &  Single photons on demand from a single molecul... &  2000 &            Nature &  Science \\   99 &   14.26 &     99 &    924 &  Collapse and revival of the matter wave field ... &  2002 &            Nature &  Science \\    99 &   13.33 &     99 &   1,092 &  Functional insights from the structure of the ... &  2000 &            Nature &  Science  \\   99 &   11.67 &     99 &    730 &  Structure of functionally activated small ribo... &  2000 &              Cell &  Biology \\   \textbf{95} &    1.79 &  \textbf{72} &     11 &  Discovery of a supernova explosion at half the... &  1998 &            Nature &  Science  \\  \textbf{90} &    1.12 &   \textbf{85} &     25 &  Generation of nonclassical motional states of ... &  1996 &  Phys. Rev. Lett. &  Science \\ \textbf{65} &    0.60 & \textbf{59} &      5 &  Direct evidence for neutrino flavor transforma... &  2002 &     AIP CONF PROC &  Science \\
	 \bottomrule
		\end{tabular}
			}
	\label{tab:my_label_nobel}
\end{table}

Moreover, it seems that ASP alleviates citation inflation towards certain types of articles. We use Computer Science as an example. Table \ref{tab:cs} lists the articles that appeared in the top 20 rankings of ASP and \citcount in the cluster. Articles in the top ASP ranking come from more concentrated topics: machine learning, control, information systems, and so on. In contrast, articles on the \citcount are led by applied papers published in interdisciplinary fields, including Physics, Biomedical Informatics, and so on, which usually receive many more citations compared to pure Computer Science articles. This may imply that ASP instead of \citcount makes for fairer comparisons when evaluating articles from different research orientations. 

\begin{table}[ht!]
	\centering
	\caption{The sets of articles with either highest 20 ASP or highest 20 \citcount in Computer Science cluster, ranked by ASP, where $n$ is the rank ASP and $k$ is the rank \citcount.}
		\resizebox{1\linewidth}{!}{
\begin{tabular}{rrrrlrll}
	\toprule
			$n$ &  ASP &  $k$ &   \citcount &                                              Title &  Year &                             Source & Cluster \\

\midrule
          1 &  768.86 &                  3 &   19,875 &                            SUPPORT-VECTOR NETWORKS &  1995 &                                       Mach. Learn. &  Computer Science \\
          2 &  613.32 &                  1 &   33,120 &                                     Random forests &  2001 &                                       Mach. Learn. &  Computer Science \\
            3 &  611.86 &                 11 &   11,543 &  Histograms of oriented gradients for human det... &  2005 &                                     PROC CVPR IEEE &  Computer Science \\
            4 &  611.01 &                  5 &   18,940 &  NavOptim coding: Supporting website navigation... &  2004 &  IEEE/WIC/ACM INTERNATIONAL CONFERENCE ON WEB I... &  Computer Science \\
            5 &  470.91 &                 10 &   14,255 &                             Long short-term memory &  1997 &                                     Neural Comput. &  Computer Science \\
            6 &  470.71 &                 21 &    7,904 &                 Wireless sensor networks: a survey &  2002 &                                      Comput. Netw. &  Computer Science \\
           7 &  456.75 &                 13 &    9,856 &                                 Bagging predictors &  1996 &                                       Mach. Learn. &  Computer Science \\
           8 &  444.63 &                  6 &   17,781 &  A fast and elitist multiobjective genetic algo... &  2002 &                          IEEE Trans. Evol. Comput. &  Computer Science \\
            9 &  423.44 &                 61 &    4,843 &                  The capacity of wireless networks &  2000 &                            IEEE Trans. Inf. Theory &  Computer Science \\
          10 &  401.27 &                 65 &    4,772 &  Space-time codes for high data rate wireless c... &  1998 &                            IEEE Trans. Inf. Theory &  Computer Science \\
           11 &  391.86 &                  9 &   14,294 &                                 Compressed sensing &  2006 &                            IEEE Trans. Inf. Theory &  Computer Science \\
           12 &  385.19 &                 16 &    8,666 &  A tutorial on Support Vector Machines for patt... &  1998 &                           Data Min. Knowl. Discov. &  Computer Science \\
        13 &  381.22 &                 64 &    4,790 &  A TRANSLATION APPROACH TO PORTABLE ONTOLOGY SP... &  1993 &                                     Knowl. Acquis. &  Computer Science \\
        14 &  374.62 &                  8 &   16,181 &  Image quality assessment: From error visibilit... &  2004 &                         IEEE Trans. Image Process. &  Computer Science \\
         15 &  366.41 &                 32 &    6,756 &  Distinctive image features from scale-invarian... &  2004 &                               Int. J. Comput. Vis. &  Computer Science \\
        16 &  364.81 &                141 &    3,034 &  Random Early Detection Gateways for Congestion... &  1993 &                              IEEE-ACM Trans. Netw. &  Computer Science \\
         17 &  361.81 &                 66 &    4,715 &               INDEXING BY LATENT SEMANTIC ANALYSIS &  1990 &                              J. Am. Soc. Inf. Sci. &  Computer Science \\
         18 &  353.45 &                  7 &   16,930 &      LIBSVM: A Library for Support Vector Machines &  2011 &                  ACM Trans. Intell. Syst. Technol. &  Computer Science \\
         19 &  346.00 &                 36 &    6,564 &  MapReduce: Simplified data processing on large... &  2004 &  USENIX ASSOCIATION PROCEEDINGS OF THE SIXTH SY... &  Computer Science \\
         20 &  341.35 &                 19 &    8,231 &  Cooperative diversity in wireless networks: Ef... &  2004 &                            IEEE Trans. Inf. Theory &  Computer Science \\
         
         \midrule
           54 &  210.11 &                  2 &   20,639 &  FAST PARALLEL ALGORITHMS FOR SHORT-RANGE MOLEC... &  1995 &                                   J. Comput. Phys. &  Computer Science \\
     95 &  141.58 &                  4 &   18,973 &     Fitting Linear Mixed-Effects Models Using lme4 &  2015 &                                    J. Stat. Softw. &  Computer Science \\

       71 &  175.51 &                 12 &   11,477 &  Research electronic data capture (REDCap)-A me... &  2009 &                                 J. Biomed. Inform. &  Computer Science \\

      63 &  191.71 &                 14 &    9,378 &  TreeView: An application to display phylogenet... &  1996 &                              Comput. Appl. Biosci. &  Computer Science \\
  22 &  315.55 &                 15 &    8,862 &            A METHOD FOR REGISTRATION OF 3-D SHAPES &  1992 &            IEEE Trans. Pattern Anal. Mach. Intell. &  Computer Science \\

  32 &  259.25 &                 17 &    8,621 &                        Generative Adversarial Nets &  2014 &                                        ADV NEUR IN &  Computer Science \\
        34 &  254.81 &                 18 &    8,540 &  Robust uncertainty principles: Exact signal re... &  2006 &                            IEEE Trans. Inf. Theory &  Computer Science \\
          111 &  132.95 &                 20 &    7,915 &  User acceptance of information technology: Tow... &  2003 &                                             MIS Q. &  Computer Science \\

\bottomrule
\end{tabular}
}
\label{tab:cs}
\end{table}

\subsubsection{Examples of other groups}
The story is different for other groups where the two metrics are not compatible. Heuristically, an article can be considered influential if it is cited by many articles, or if it, though not directly cited by many, is cited by other influential article(s) with many citations. As mentioned, \citcount counts only direct citations, while ASP is able to reflect the prestige including these indirect citations. An Engineering article “Blind decorrelation and deconvolution algorithm for multiple-input multiple-output system: I. Theorem derivation” was ranked as high in the $99^\text{th}$ percentile (top 1\%) with ASP value (21.75) but relatively low in the $23^\text{th}$ percentile by \citcount (with 1 count). The Computer Science article “A computer algebra system based on order-sorted algebra” was ranked in the $99^\text{th}$ percentile by the ASP (19.94) but in the $30^\text{th}$ percentile by \citcount (with 2 counts). The Social Science article “Vision and the autonomous symbol in the works of Lorrain, Jean-stage sets and obstacles” was ranked in the $99^\text{th}$ percentile by the ASP (17.34), but in the $23^\text{th}$ percentile by \citcount (with 1 count). Although the above may be argued as special cases, there are more examples of articles with small \citcount but high ASP. Specifically, 43.83\% of articles have high ASP values and a small number of citation counts. In contrast, there are 33.62\% articles with low ASP but high \citcount. 


\subsection{ASP and Journal Grade}
\label{sec:asp_journal}
Historically, journals are used to publishing (printing and distributing) scientific results but less so in this millennium. One can find important publications in \textit{arXiv.org} even though they are not published in known (refereed) journals. Journals are also used to cluster/sort/filter articles regarding particular topics. If one looks into “Applied Statistics,” articles in the journals, one expects articles to be about applied statistics. Over the years, there might have been a shift in the topics, so the journal name might not exactly match its contents anymore. Important journals (by the usual metrics) like \textit{Science} or \textit{Nature} are not sorted by topic at all. Since the mid-20th century, under the hypothesis that the more citations an entity (e.g., a scholar, an institute, or a journal) receives, the higher scientific impact it has, citation-based metrics have been developed and became popular, particularly for evaluating a journal’s scientific prestige at the journal level, where the total citation counts of all articles published in the journal are considered over a certain time period. See, for example, Science Citation Index (SCI), CiteScore, Impact Factor (IF), Hirsch's bibliometrics index (H-Index), and SCImago Journal Rank (SJR). Given the publicly available journal-level citation metrics, it becomes common and recognized to judge an article’s scientific value based on which journal publishes it. 

\begin{figure}[ht!]
	\centering
	\includegraphics[width=0.58\textwidth, height=0.3\textheight]{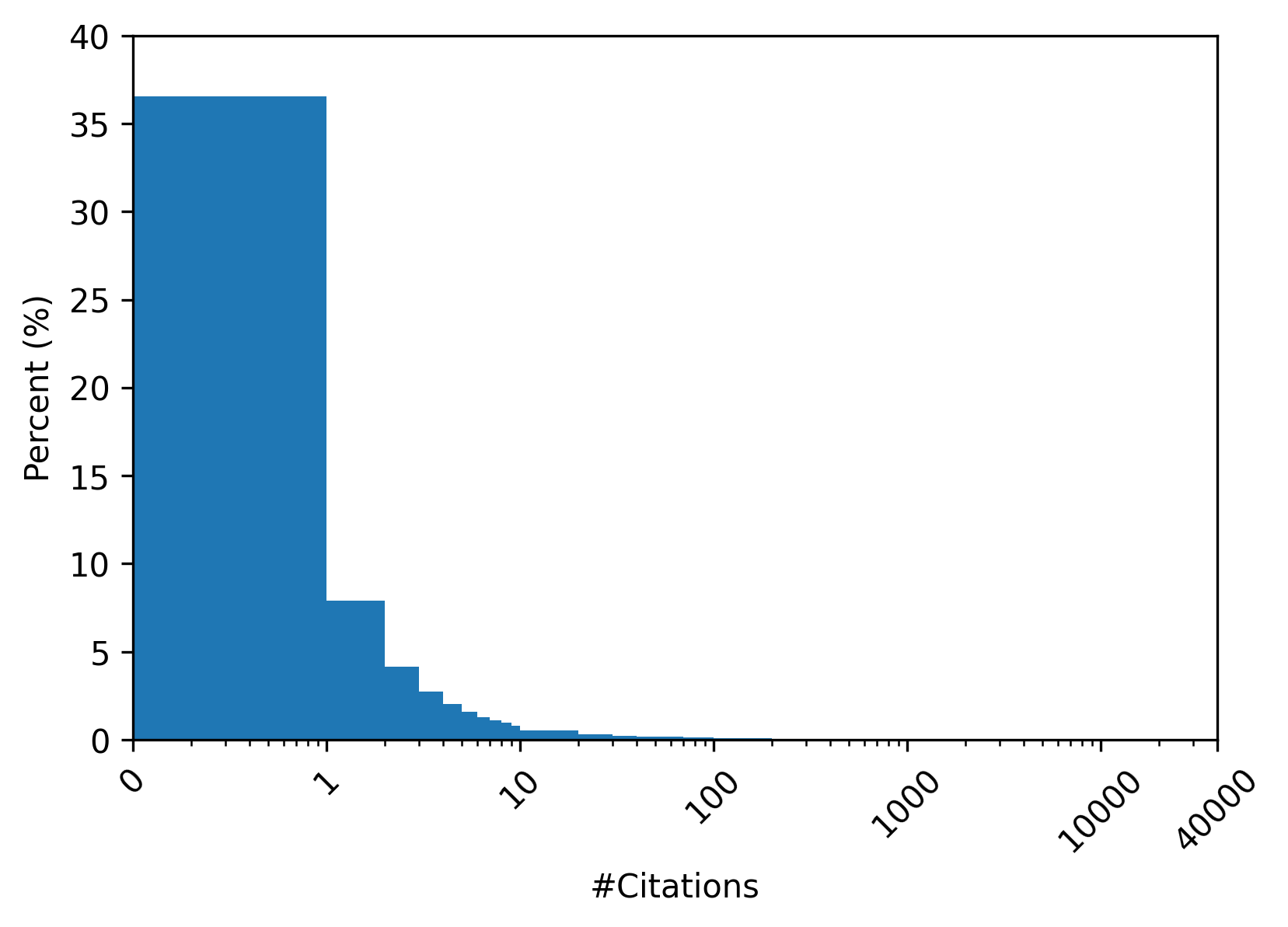}
	\caption{Histogram for \citcount for articles published between 1981 and 2020 in \textit{Nature} (ISSN: 0028-0836, 1476-4687) }
	\label{fig:Math_journal}
\end{figure}

Admittedly, an article published in a highly regarded journal is more likely to be read and cited, increasing its chances of becoming “important” and influential in scientific society. However, it is often misleading to evaluate an article’s scientific prestige indirectly based on a journal’s rank. The quality of an article is unlikely to change with a journal’s impact. Instead, a journal’s value will be improved if it publishes important articles. Meanwhile, the distribution of \citcount is right-skewed with a long tail slowly fragmenting towards the extremely large citations, implying that the majority of papers published in the top tiered journal are overvalued when judging with journal prestige. Figure \ref{fig:Math_journal} presents the empirical density of \citcount in \textit{Nature}  (ISSN: 0028-0836, 1476-4687) from 1981 to 2020. The journal issued by Nature Research is a
 prestigious journal in multidisciplinary science and has been well recognized by the journal-level citation metrics, with an IF of 42.778 in 2019, H-index of 1226, and SJR of 15.993 in 2020. According to journal grade information, its max, min, median of \citcount are 3157, 1, and 4, respectively and ASP are 60.68, 0.5, and 0.5, respectively. Given  that \textit{Nature} is considered as one of the most prestigious journals, all the papers should have higher chances of being cited. Nevertheless, among the  121,107 articles published in the journal obtained from the WoS data between 1981 and 2020, 36.56\% were never cited, yet all would be considered top publications if journal grade is used as an evaluation metric.

Simultaneously, an important article may be undervalued if it is not published in a prestigious journal. Some essential works have been known, to introduce innovations that are too advanced, were rejected by conservative referees and published in less prestigious journals. A famous example is the article “The market of lemons,” written by George Akerlof in 1966 which won the 2001 Nobel Memorial Prize in Economic Sciences. The article was rejected by three renowned journals and was finally accepted and published 
upon the fourth submission in 1978 in \textit{The Quarterly Journal of Economics}.

\begin{table}[ht]
\centering
\caption{Journal grade: ASP and \citcount}\label{tab:SJR_cite_tab:SJR_ASP}
{\tiny
\begin{tabular}{ll|r@{; }lr|r@{; }lr|r@{; }lr|r@{; }lr} 
\toprule

\multirow{2}*{} & \multirow{2}*{} & \multicolumn{6}{c}{SJR} & \multicolumn{6}{c}{H-index} \\ \cline{3-14}
&& \multicolumn{3}{c}{ASP} & \multicolumn{3}{c|}{\citcount} & \multicolumn{3}{c}{ASP} & \multicolumn{3}{c}{\citcount}\\
\cline{3-14}
&& \multicolumn{2}{c}{range} & \multicolumn{1}{c|}{mean} &
\multicolumn{2}{c}{range} & \multicolumn{1}{c|}{mean} &
\multicolumn{2}{c}{range} & \multicolumn{1}{c|}{mean} &
\multicolumn{2}{c}{range} & \multicolumn{1}{c}{mean} \\
\midrule
\multirow{4}*{min} & Q1/H1 &[0.50 & 0.70] & (0.50) & 
[0 & 13] & (0.02)  & 
[0.50 & 0.70] & (0.50) & 
[0 & 13] & (0.01)\\
                   & Q2/H2 &[0.50 & 0.52] & (0.50) & 
                   [0 & 1] & (0.01) & 
                   [0.50 & 0.55] & (0.50) 
                   & [0 & 5] & (0.01)\\
                   & Q3/H3 &[0.50 &0.55] &(0.50) &
                   [0 &2] &(0.01)  & 
                   [0.50 &0.59] &(0.50)  &
                   [0 &11] &(0.01)\\
                   & Q4/H4 &[0.50 &0.67] &(0.50) & 
                   [0 &28] &(0.04)  & 
                   [0.50 &0.67] &(0.50)  &
                   [0 &28] &(0.09)\\
\midrule
\multirow{4}*{mean} & Q1/H1 &[0.50 &7.95] &(0.90) & 
[0.05 &529.58] &(25.85)  & 
[0.51 &7.95] &(0.88) & 
[0.37 &529.58] &(24.21)\\
                   & Q2/H2 &[0.50 &6.19] &(0.69) & 
                   [0.02& 103.59] &(11.19) & 
                   [0.50& 1.73] &(0.62)  & 
                   [0.07 &65.31] &(6.26)\\
                   & Q3/H3 &[0.50 &2.23] &(0.61) &  
                   [0.01 &37.41] &(6.01)  & 
                   [0.50 &0.93] &(0.55)  & 
                   [0.04 &73.18] &(1.54)\\
                   & Q4/H4 &[0.50 &0.89] &(0.56) 
                   &  [0.01 &34.16] &(2.66)  & 
                   [0.50 &0.70] &(0.52)  &
                   [0.01 &28] &(0.41)\\                   
\midrule
\multirow{4}*{median} & Q1/H1 &[0.50 &3.98] &(0.66) & 
[0 &330.5] &(13.19)  & 
[0.50 &3.98] &(0.66) & 
[0 &330.5] &(12.43)\\
                   & Q2/H2 &[0.50 &1.65] &(0.58) & 
                   [0 &32.5] &(5.93) & 
                   [0.50 &1.01] &(0.54) 
                   & [0.5 &27.5] &(3.26)\\
                   & Q3/H3 &[0.50 &1.02] &(0.54) &  
                   [0 &23] &(3.05)  & 
                   [0.50 &0.85] &(0.51)  & 
                   [0.5 &46] &(0.59)\\
                   & Q4/H4 &[0.50 &0.67] &(0.51) &  
                   [0& 28] &(1.19)  & 
                   [0.50 &0.67] &(0.50)  & 
                   [0.5 &28] &(0.13)\\ 
\midrule
\multirow{4}*{max} & Q1/H1 &[0.60 &1570.09] &(23.75) & 
[3 &58,002] &(1204.35)  &
[0.62 &1570.09] &(22.69) & 
[9 &58,002] &(1139.18)\\
                   & Q2/H2 &[0.56& 307.57] &(9.10) & 
                   [1 &26,895] &(418.78) & 
                   [0.52 &163.92] &(3.67)  & 
                   [2 &16,693] &(133.74)\\
                   & Q3/H3 &[0.54 &220.06] &(5.78) & 
                   [1 &16,693] &(224.73)  & 
                   [0.56 &14.02] &(1.91)  & 
                   [1 &341] &(27.34)\\
                   & Q4/H4 &[0.51 &56.61] &(3.05) & 
                   [1 &3,432] &(92.26)  & 
                   [0.51 &6.82] &(1.45)  & 
                   [1 &90] &(11.50)\\ 
\midrule
\multirow{3}*{Remark} &  & \multicolumn{6}{c|}{number of journals} & \multicolumn{6}{c}{number of journals}\\
                      &  & \multicolumn{6}{c|}{Q1: 5,117, Q2: 3,242,}  & \multicolumn{6}{c}{H1: 6,621, H2: 3,123,}  \\
                      & &\multicolumn{6}{c|}{Q3: 1,791, Q4: 664} & \multicolumn{6}{c}{H3: 792, H4: 313} \\

\bottomrule
\end{tabular}}
\end{table}

We considered the 65,045 journals in the WoS data and categorized these journals according to the SJR ( Scientific Journal Ranking) in 2020 and the H-index\footnote{We followed SJR ranking to separate journals into 4 groups: Q1 to Q4. We separated the H-index according to quartile, leading to 4 groups labelled as H1 to H4.} respectively. When merging the data according to ISSN, 35,952 journals had complete information, among which 10,814 journals have both citation and ASP records. We then conducted a statistical analysis to investigate the relation between ASP and journal grade. Table \ref{tab:SJR_cite_tab:SJR_ASP} presents the range of mean, median, min, and max of ASP and \citcount of journals according to the journal grade. It shows that, in terms of average value, ASP and \citcount, when aggregated to journal level, are consistent to the SJR journal grade. However, the minimum average of \citcount increases from Q1 to Q4 level journals, meaning that \citcount is not aligned to journal grade, where Q4 journals supply the minimum average of 0.04.
Moreover, the maximum of min \citcount a journal can get shows that some Q4 level journals are superior with 28 citations, while the Q1 level journal has only 13 cites. 
The inconsistency between \citcount and journal grade disappears when the H-index is used. This is no surprise as the H-index essentially reflects the same information as the total number of citations according to probability theory, see \cite{Krattenthaler+2021+124+128}.

\begin{figure}[ht!]
    \centering
    \begin{subfigure}[b]{0.5\textwidth}
        \centering
        \includegraphics[width=1\linewidth, height=0.3\textheight]{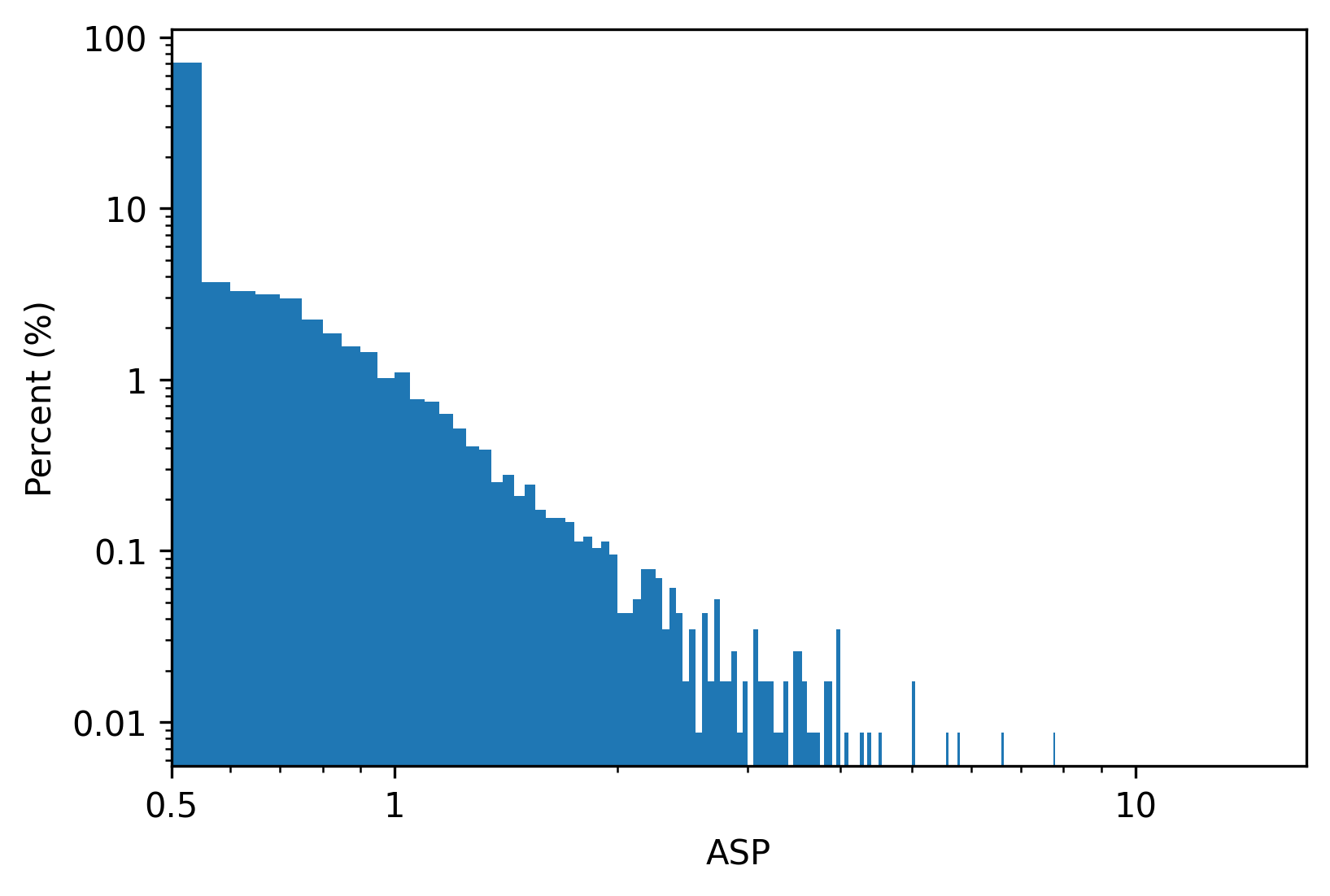}
        \caption{\textit{Molecular Therapy}}
        
    \end{subfigure}%
    \begin{subfigure}[b]{0.5\textwidth}
        \centering
        \includegraphics[width=1\linewidth, height=0.3\textheight]{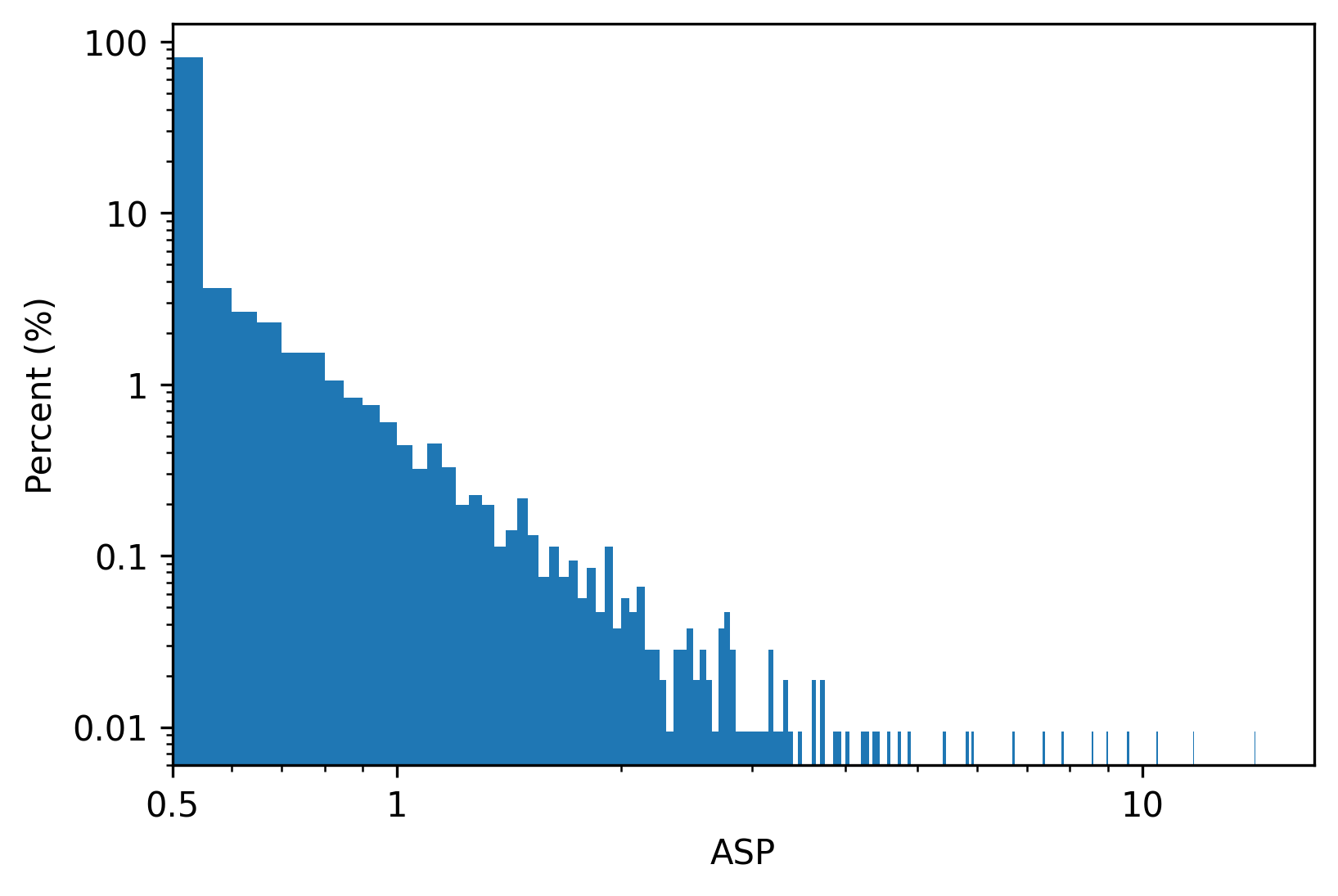}
        \caption{\textit{Pharmacoepidemiology and Drug Safety}}
        
    \end{subfigure}
     \begin{subfigure}[b]{0.5\textwidth}
        \centering
        \includegraphics[width=1\linewidth, height=0.3\textheight]{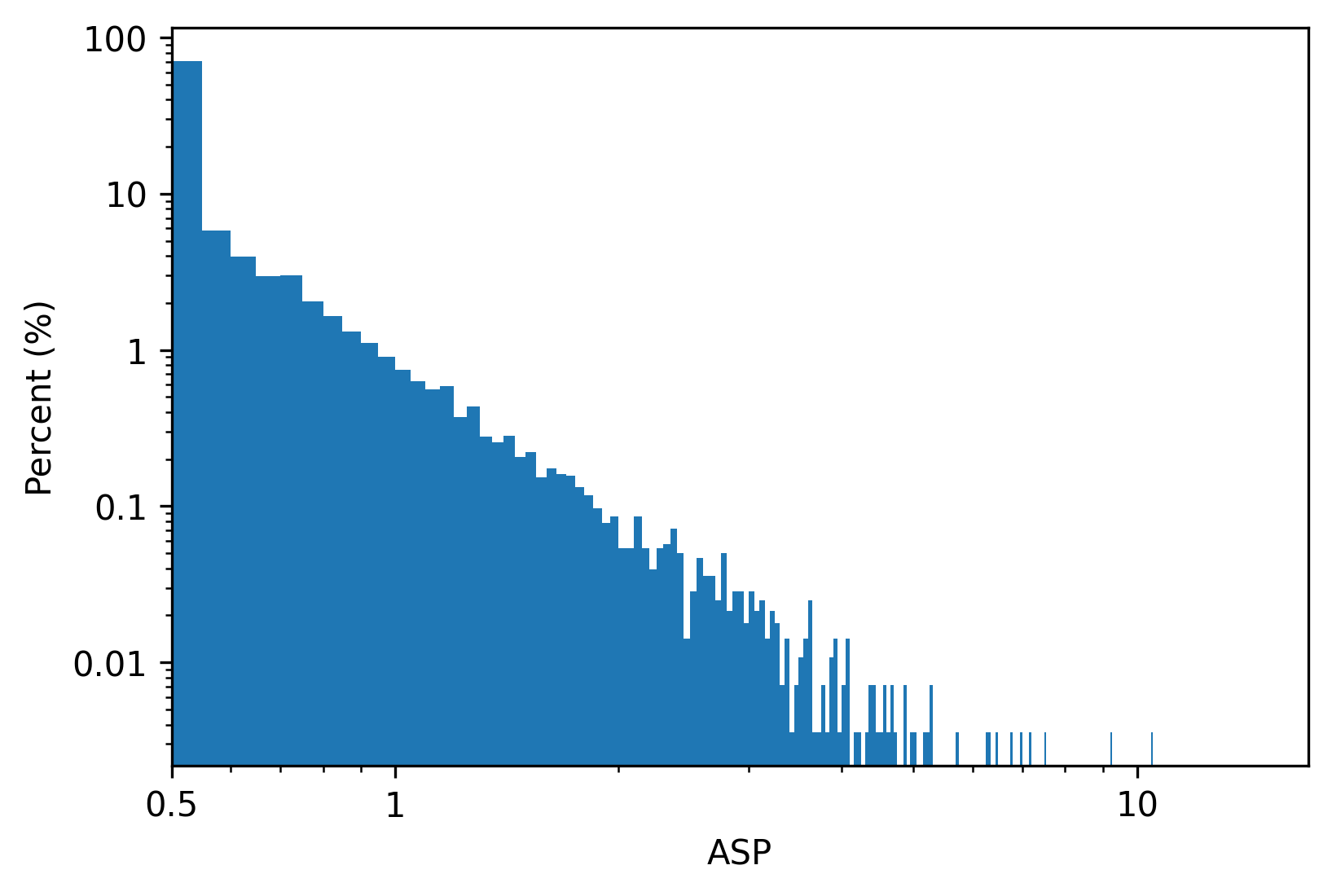}
        \caption{\textit{Veterinary Record}}
       
    \end{subfigure}%
    \begin{subfigure}[b]{0.5\textwidth}
        \centering
        \includegraphics[width=1\linewidth, height=0.3\textheight]{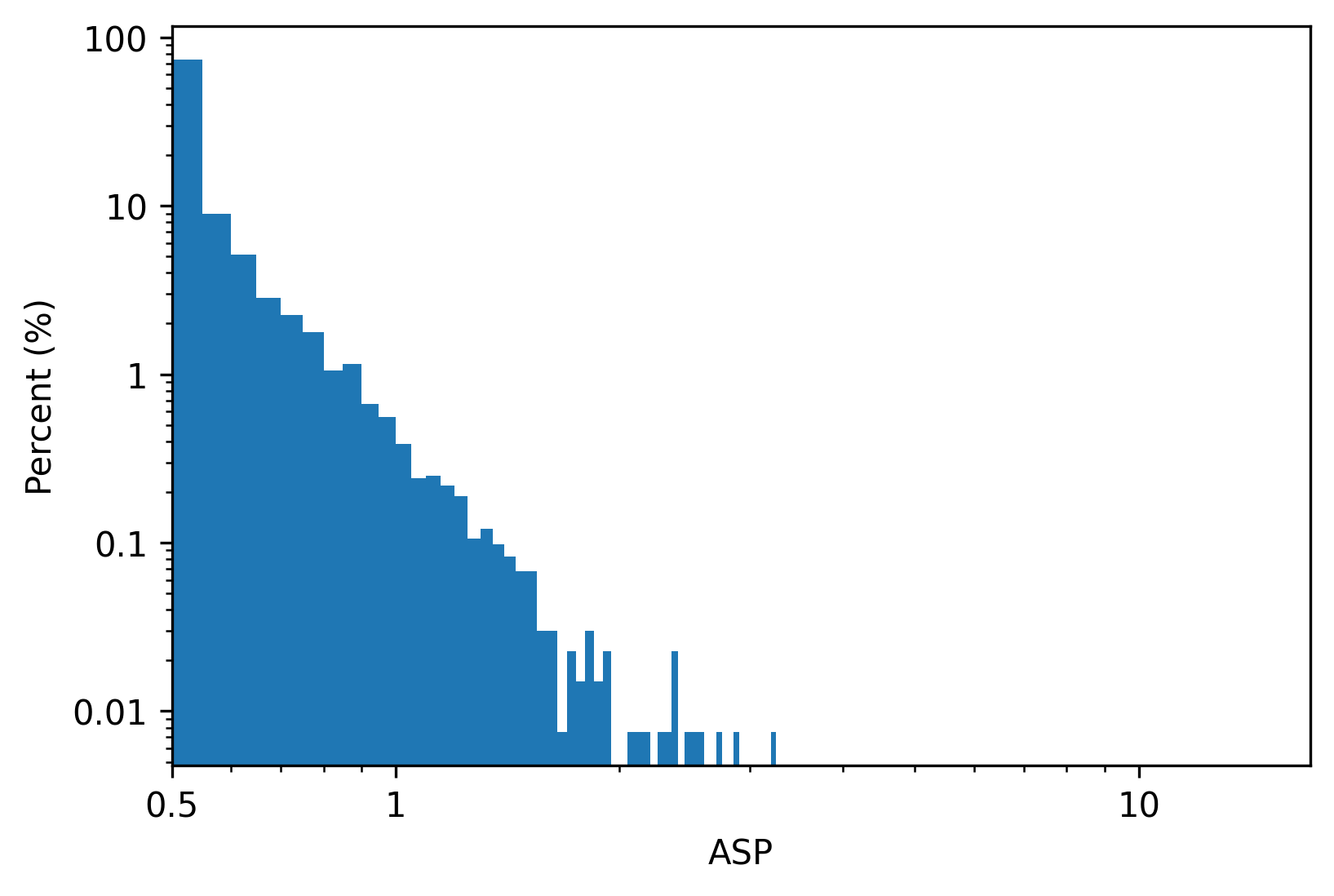}
        \caption{\textit{Deutsche Medizinische Wochenschrift}}
    
    \end{subfigure}
\caption{ASP distribution for Medicine articles published in four journals in different quartile of $\text{SJR}_{2020}$ ranking. Articles are published between 1990 and 2015}
 \label{sjr:journal}
\end{figure}

Nevertheless, we argue that journal grade is not the right metric to evaluate an article's scientific prestige. Figure \ref{sjr:journal} demonstrates the distribution of ASP in 4 journals in the cluster of Medicine with different SCImago Journal Ranks (SJR) by \cite{gonzalez2010new}. The histograms show the ASP distribution in Medicine journals, namely \textit{Molecular Therapy} with $\text{SJR}_{2020}=3.871,$ \textit{Pharmacoepidemiology and Drug Safety} with $\text{SJR}_{2020}=1.023,$ \textit{Veterinary Record} with $\text{SJR}_{2020}=0.261,$ and \textit{Deutsche Medizinische Wochenschrift} with $\text{SJR}_{2020}=0.151.$  It shows that the ASP follows logarithmic law distribution regardless of the journal grade, where 60.74\%, 72.59\%, 61.43\% and 50.60\%, respectively, are not cited. This means that no matter where an article is published, there is a chance of no scientific influence. For the articles published in grade I journals, this is more questionable given that the high journal grade enhances the visibility of articles. In short,  an article should not be judged solely based on the grade of the journal it is published in.

\subsection{References and Coauthors}
\label{sec:asp_na_nref}
\begin{figure}[ht!]
\begin{subfigure}[b]{0.5\textwidth}
		\centering
		\includegraphics[width=1.1\textwidth, height=0.3\textheight]{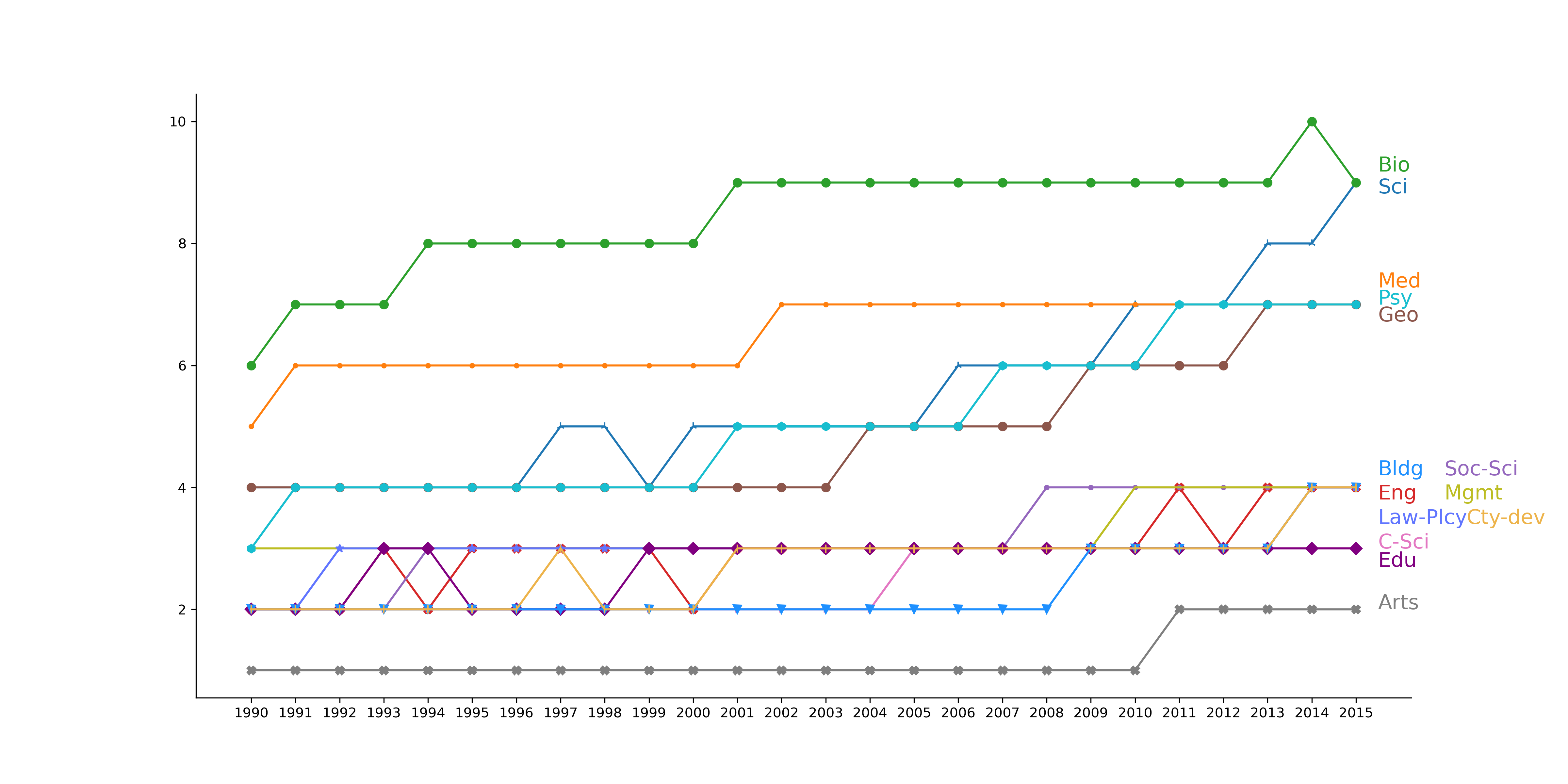}	\caption{References}	
		\label{fig:References_years}
	    \end{subfigure} 
	    \begin{subfigure}[b]{0.5\textwidth}
		\centering
		\includegraphics[trim=0 0.3cm 0 2.3cm, clip, width=1.1\textwidth, height=0.3\textheight]{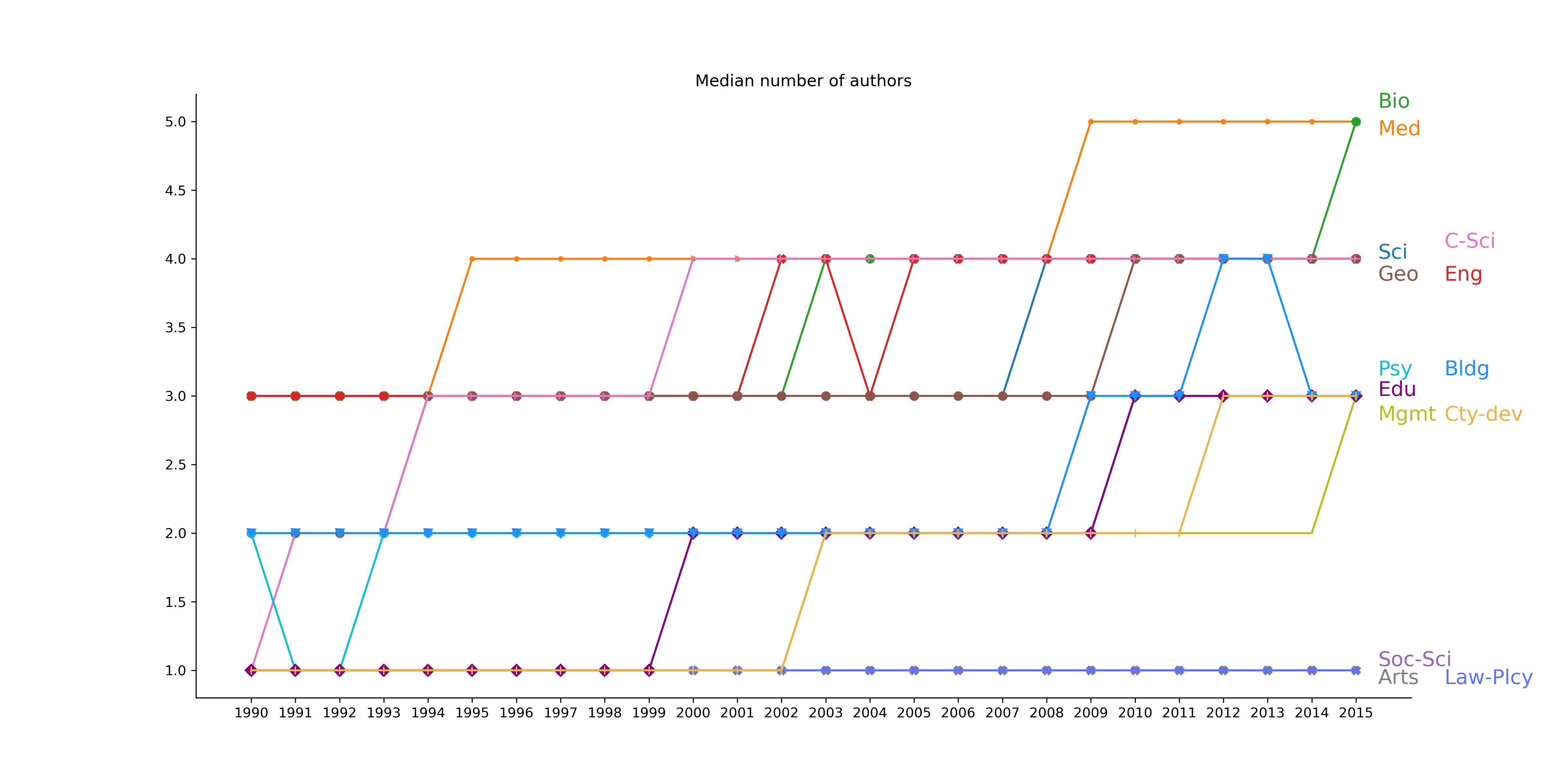} 
		\caption{Coauthors}	
		\label{fig:Coauthors_years}

   \end{subfigure} 
	\caption{Medians of References and Coauthors per article for 14 clusters between 1990 and 2015.}
	\label{fig_stat:Median_ref}
\end{figure}

The WoS data shows that most articles (75\%) are coauthored by no more than five authors and 75\% of articles have no more than 36 references. Figure \ref{fig_stat:Median_ref} presents the median of references and coauthors per article for the articles published from 1990-2015. In terms of references, Biology leads with 50\% of articles referring to 6-10 previous articles, followed by Medicine with 5-7 references. In general, there is a mild increase of references per article in almost all the clusters, yet at different rates. Science exhibits a dramatic increase from four references per article in 1990 to nine in 2015. Geography and Psychology display significant increases too, though at a slower speed. Arts, Education, and Law \& Policy have the smallest number of references, though a sharp increase  has occurred since 2010. Computer Science appears at the bottom, possibly because conference proceedings rather than articles are more recognized in this cluster. 

Regarding the number of coauthors, Medicine and Biology on average involve a bigger team, and the median increases over time and has reached five coauthors in recent years. On the other hand, Social Science, Arts, and Law \& Policy have smaller size, where at least 50\% of articles are sole-authored. Science, Computer Science, Engineering, and Geography have had increasingly more coauthors over the recent years. The median  number of authors in Computer Science increased from one in 1990 to four in 2015 on median. The trend is not always monotonic in other clusters. There is, for example, a continuous increase in in the number of coauthors in Building from 2008-2012, followed by a sudden drop from 2013-2015. 

\begin{figure}[ht!]
    \centering
    \begin{subfigure}[b]{0.5\textwidth}
        \centering
        \includegraphics[width=1\linewidth, height=0.3\textheight]{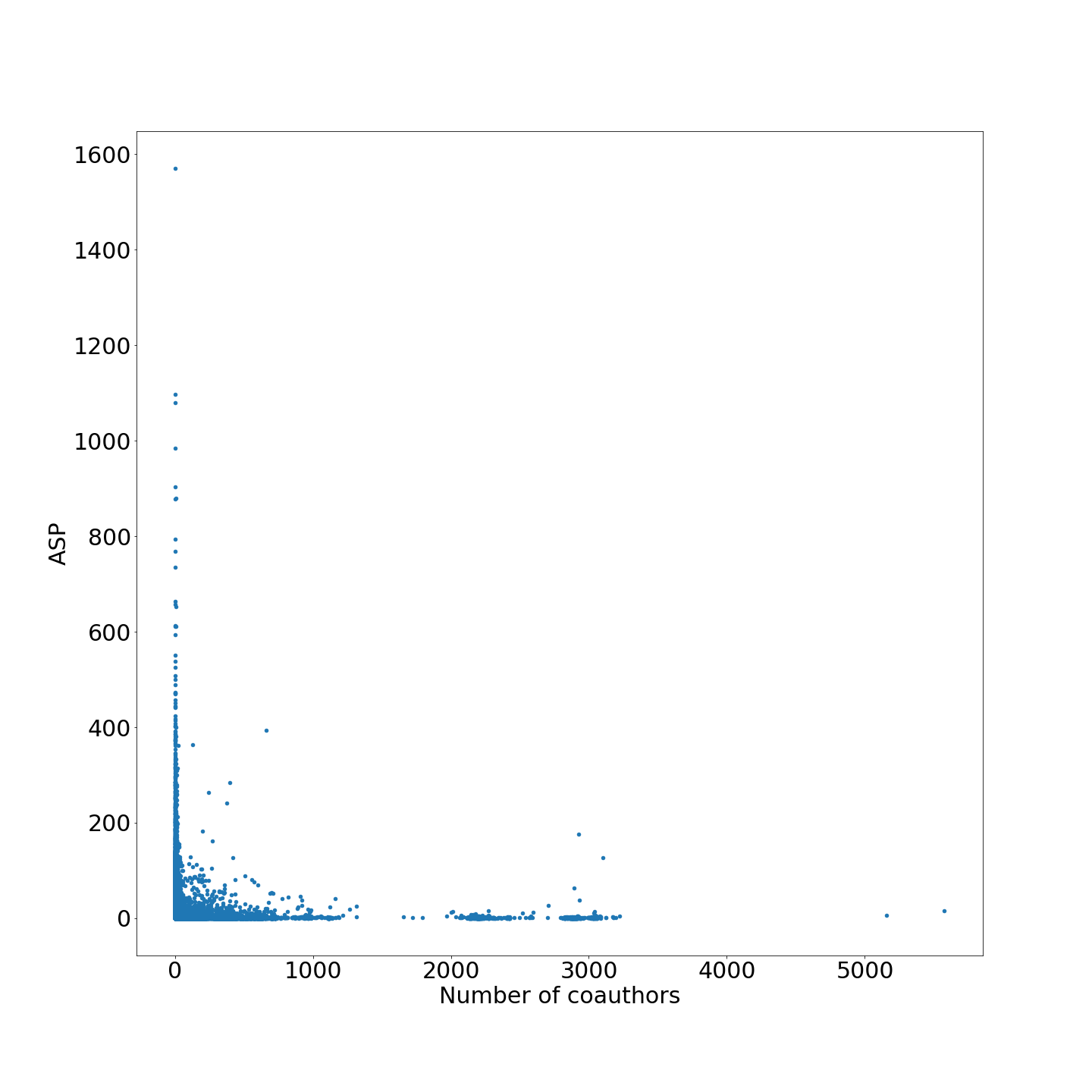}
        \caption{ASP vs number of coauthors.}
        \label{fig:ASP_NA}
    \end{subfigure}%
    \begin{subfigure}[b]{0.5\textwidth}
        \centering
        \includegraphics[width=1\linewidth, height=0.3\textheight]{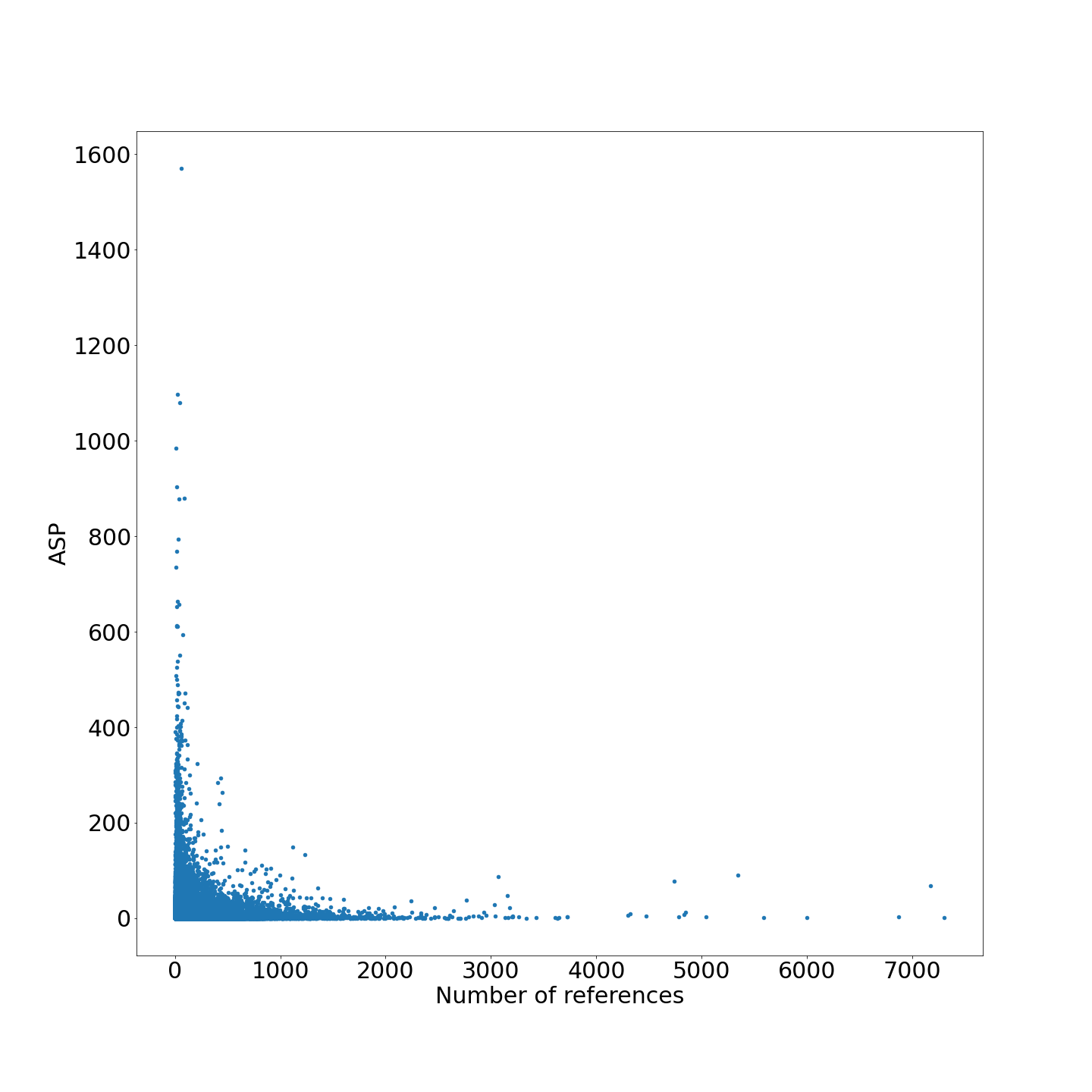}
        \caption{ASP vs number of references.}
       \label{fig:ASP_NRef}
    \end{subfigure}
\caption{ ASP vs number of coauthors and references}
 \label{fig:ASP_NA_NRef}
\end{figure}

To what extent is references or number of coauthors helpful to improve ASP? Figure \ref{fig:ASP_NA_NRef} presents the scatter plot of the ASP versus the number of coauthors (panel \subref{fig:ASP_NA}) and references (panel \subref{fig:ASP_NRef}). We find there is no evidence that more references or more coauthors improve ASP.

\section{Conclusion}
\label{sec:conc}
We analyzed a large-scale WoS citation network with millions of articles from 254 subjects published between 1981 and 2020. We proposed the ASP index to evaluate the scientific importance of individual articles in the network using the eigenvector centrality metric. We found that there is a high correlation between the ASP and the \citcount among the top 10\% of articles but a significantly minor dependence for the rest articles. There is little evidence of influence of the number of references and coauthors on the article's scientific quality. Furthermore, ASP minimizes the difference in scientific quality distribution among the disciplines. In consistent to the fact that the quality distribution of journal articles is dramatically right-skewed, the articles' scientific prestige should not be judged based on the journal grades, which is supported by our analysis. With a parallel algorithm on sparse data-structures, we can obtain the ASPs for over 30 million articles in a few seconds, demonstrating that it is computationally feasible to evaluate all articles individually. Without question, there is still room for improvement in evaluating the prestige and impact of scientific articles. Our analysis showcases that there is no computational hurdle from including further aspects. For example, the increasing use of unique and well defined IDs like OrcID will allow in the future a reliable evaluation of author/co-authorship relations over multiple articles and citations. Meanwhile, the quality of the data is of crucial importance. We noticed that it seems very likely that a considerable number of references is missing from WoS, though it corresponds to small percentage in the large-scale citation network. Publications channels continue to expand, the importance of Proceedings and Open Access repositories e.g. \textit{arXiv.org}, or self publishing via Social Media like ResearchGate is constantly increasing. Maybe it is time to end judging a publication by where it is published but to compute individually how much ``prestige'' it manages to attract. As an additional benefit this would make the introduction of new publication outlets much easier. 

\section{Acknowledgement}
The work for this article has been conducted in the Research Campus MODAL funded by the German Federal Ministry of Education and Research (BMBF) (fund numbers 05M14ZAM, 05M20ZBM).

\begin{appendices}
\section{}
\subsection{Sample of citation data} 
\label{app1}

Figure  \ref{tab:variables} illustrates the raw information of an article entitled ``Basic local alignment search tool'' by Altschul Stephen F. and Gish Warren and others, published in 1990 in the \textit{Journal of Molecular Biology} which received the most citations, i.e. 58,002 citations in our data.
\renewcommand{\thefigure}{A.\arabic{figure}}
\setcounter{figure}{0}   
\begin{figure}[H]
		\includegraphics[width=1.1\textwidth, height=0.8\textheight]{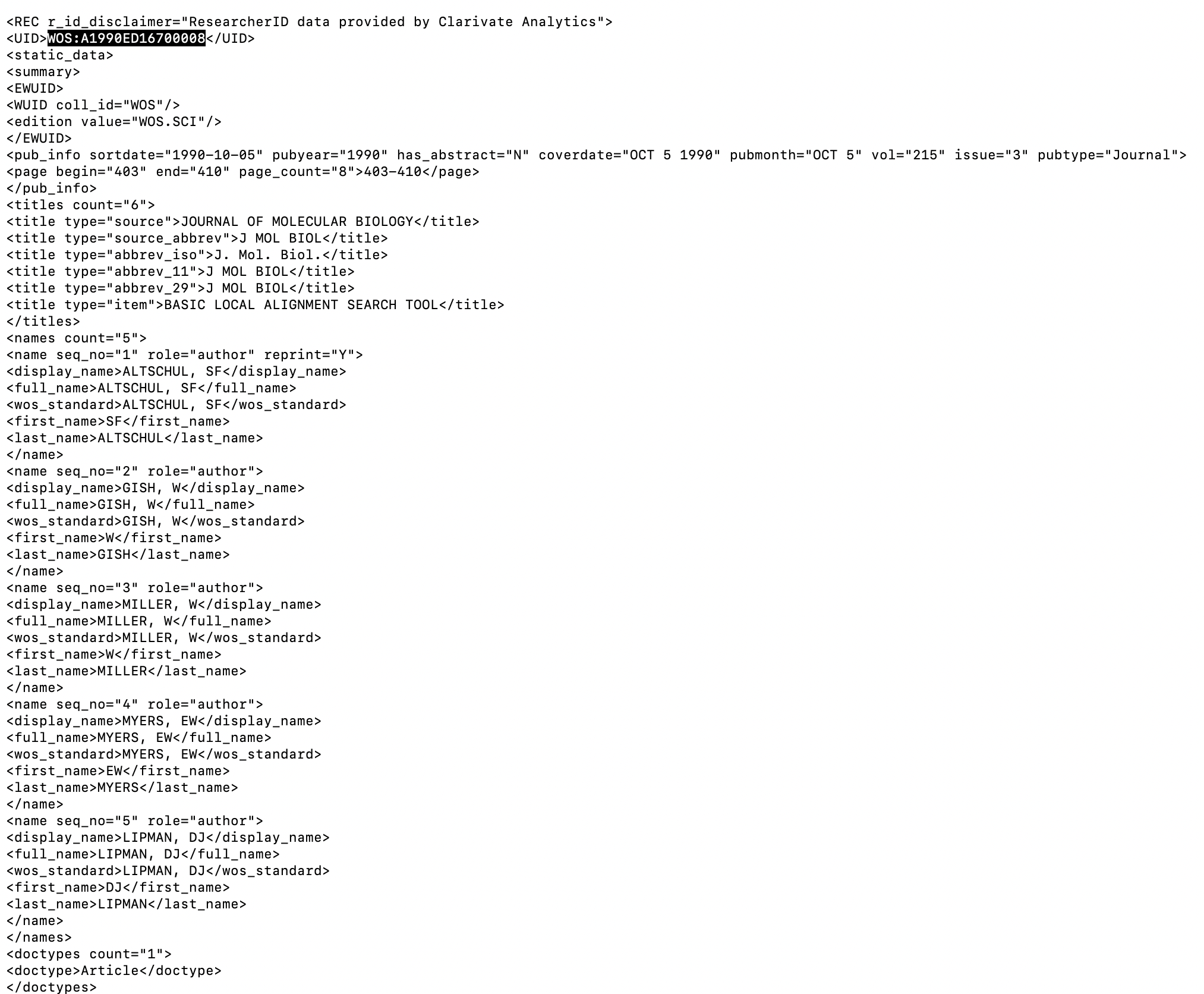}
	\caption{Sample of the Web of Science dataset for the most cited article entitled "Basic local alignment search tool" published in 1990 in \textit{Journal of Molecular Biology} with the UID "WOS:A1990ED16700008" and 5 coauthors.}
	\label{tab:variables}
\end{figure}

\subsection{The 14 clusters and 254 subjects} 
\label{app2}
We group the disciplines into 14 scientific clusters where the cluster information is summarised in Table \ref{tab:clusters}. Science and Medicine form the two biggest groups, by 43 and 56 disciplines respectively. The two clusters also have the largest number of articles with more than 8.7 millions for each. This is almost double of the 3rd largest cluster, Biology. In terms of citations, the 3rd largest cluster Biology stands out with a median of 16, while Medicine and Geography, the second best, have 11 citations on average, which are 5 citations less than Biology. It is also interesting to note that Psychology, though with much less articles published (0.6 million articles), and relatively smaller citations (8 each article), has the largest number of references (30 per article). Among the 14 clusters, Social Science and Arts have 0 median citation, which means 50\% of articles in the two clusters are never cited. The two clusters also have the smallest number of references (1 per article).

\setcounter{table}{0}
\renewcommand{\thetable}{A.\arabic{table}}
\begin{table}[ht!]
  \centering
 \caption{Classification of the 254 disciplines into 14 clusters (1990 and 2015).}
   \label{tab:clusters}%
  \tiny 
    \begin{tabular}{L{1.2cm}|L{10.8cm}|R{0.6cm}|R{1.2cm}|R{0.7cm}|R{0.8cm}}
 	\hline
 	\hline
 	\textbf{Cluster}  & \textbf{Disciplines}& \textbf{Sub.} & \textbf{Articles} & \textbf{Refs} & \textbf{\citcount} \\
	\hline
	 Science          &    Acoustics; Astronomy \& Astrophysics; Chemistry, Analytical; Chemistry, Applied; Chemistry, Inorganic \& Nuclear; Chemistry, Medicinal; Chemistry, Multidisciplinary; Chemistry, Organic; Chemistry, Physical; Crystallography; Electrochemistry; Engineering, Chemical; Imaging Science \& Photographic Technology; Materials Science, Biomaterials; Materials Science, Ceramics; Materials Science, Characterization \& Testing; Materials Science, Coatings \& Films; Materials Science, Composites; Materials Science, Multidisciplinary; Materials Science, Paper \& Wood; Materials Science, Textiles; Mathematics; Mathematics, Applied; Mathematics, Interdisciplinary Applications; Mechanics; Multidisciplinary Sciences; Nanoscience \& Nanotechnology; Nuclear Science \& Technology; Optics; Physics, Applied; Physics, Atomic, Molecular \& Chemical; Physics, Condensed Matter; Physics, Fluids \& Plasmas; Physics, Mathematical; Physics, Multidisciplinary; Physics, Nuclear; Physics, Particles \& Fields; Polymer Science; Spectroscopy; Statistics \& Probability; Thermodynamics; Quantum Science \& Technology; Green \& Sustainable Science \& Technology & 43 & 8,790,911 &     21 &       9       \\
    \hline
    Medicine         &   Allergy; Anatomy \& Morphology; Andrology; Anesthesiology; Audiology \& Speech-Language Pathology; Cardiac \& Cardiovascular Systems; Clinical Neurology; Critical Care Medicine; Dentistry, Oral Surgery \& Medicine; Dermatology; Emergency Medicine; Endocrinology \& Metabolism; Ethics; Gastroenterology \& Hepatology; Genetics \& Heredity; Geriatrics \& Gerontology; Health Care Sciences \& Services; Health Policy \& Services; Hematology; Immunology; Infectious Diseases; Integrative \& Complementary Medicine; Medical Ethics; Medical Informatics; Medical Laboratory Technology; Medicine, General \& Internal; Medicine, Research \& Experimental; Microscopy; Neuroimaging; Neurosciences; Nursing; Obstetrics \& Gynecology; Oncology; Ophthalmology; Orthopedics; Otorhinolaryngology; Pathology; Pediatrics; Peripheral Vascular Disease; Pharmacology \& Pharmacy; Physiology; Primary Health Care; Psychiatry; Public, Environmental \& Occupational Health; Radiology, Nuclear Medicine \& Medical Imaging; Rehabilitation; Respiratory System; Rheumatology; Sport Sciences; Surgery; Toxicology; Transplantation; Tropical Medicine; Urology \& Nephrology; Veterinary Sciences; Virology &             56 &  8,717,937 &     24 &      11 \\
    \hline
Biology         &  Biochemical Research Methods; Biochemistry \& Molecular Biology; Biodiversity Conservation; Biology; Biophysics; Biotechnology \& Applied Microbiology; Cell Biology; Cell \& Tissue Engineering; Developmental Biology; Ecology; Entomology; Evolutionary Biology; Food Science \& Technology; Horticulture; Limnology; Marine \& Freshwater Biology; Mathematical \& Computational Biology; Microbiology; Mycology; Nutrition \& Dietetics; Oceanography; Ornithology; Parasitology; Plant Sciences; Reproductive Biology; Soil Science; Zoology  &             27  &  3,900,425 &   33 &      16 \\
\hline
Engineering       &  Automation \& Control Systems; Energy \& Fuels; Engineering, Aerospace; Engineering, Biomedical; Engineering, Electrical \& Electronic; Engineering, Environmental; Engineering, Industrial; Engineering, Manufacturing; Engineering, Marine; Engineering, Mechanical; Engineering, Multidisciplinary; Engineering, Ocean; Engineering, Petroleum; Ergonomics; Instruments \& Instrumentation; Metallurgy \& Metallurgical Engineering; Remote Sensing; Robotics; Telecommunications  &             19 &  2,873,037 &     13 &       2\\
\hline
Social Science   &  Anthropology; Area Studies; Behavioral Sciences; Communication; Criminology \& Penology; Demography; Ethnic Studies; Family Studies; Gerontology; History; History Of Social Sciences; Hospitality, Leisure, Sport \& Tourism; Humanities, Multidisciplinary; Information Science \& Library Science; Philosophy; Religion; Social Issues; Social Sciences, Biomedical; Social Sciences, Interdisciplinary; Social Sciences, Mathematical Methods; Social Work; Sociology; Substance Abuse; Women's Studies  &             24 &  1,937,497 &     1 &       0 \\
\hline
Geography         &  Agricultural Engineering; Agriculture, Dairy \& Animal Science; Agriculture, Multidisciplinary; Agronomy; Engineering, Geological; Environmental Sciences; Environmental Studies; Fisheries; Forestry; Geochemistry \& Geophysics; Geography; Geography, Physical; Geology; Geosciences, Multidisciplinary; Meteorology \& Atmospheric Sciences; Mineralogy; Mining \& Mineral Processing; Paleontology; Water Resources   &             19 &  1,761,068 &     28 &      11\\
\hline
Computer Science  &  Computer Science, Artificial Intelligence; Computer Science, Cybernetics; Computer Science, Hardware \& Architecture; Computer Science, Information Systems; Computer Science, Interdisciplinary Applications; Computer Science, Software Engineering; Computer Science, Theory \& Methods  &              7 &  1,682,506 &    15 &       2\\
\hline
Arts              &  Archaeology; Art; Asian Studies; Classics; Cultural Studies; Dance; Film, Radio, Television; Folklore; Language \& Linguistics; Linguistics; Literary Reviews; Literary Theory \& Criticism; Literature; Literature, African, Australian, Canadian; Literature, American; Literature, British Isles; Literature, German, Dutch, Scandinavian; Literature, Romance; Literature, Slavic; Logic; Medieval \& Renaissance Studies; Music; Poetry; Theater  &             24 &  1,276,883 &     1 &       0\\
\hline
Management        &  Business; Business, Finance; Economics; Management; Operations Research \& Management Science; Public Administration &              6 &   623,059 &     22 &       4 \\
\hline
Psychology       & History \& Philosophy Of Science; Psychology; Psychology, Applied; Psychology, Biological; Psychology, Clinical; Psychology, Developmental; Psychology, Educational; Psychology, Experimental; Psychology, Mathematical; Psychology, Multidisciplinary; Psychology, Psychoanalysis; Psychology, Social &             12 &   614,806 &     30 &       8\\
\hline
Law and Policy   &  Agricultural Economics \& Policy; Industrial Relations \& Labor; International Relations; Law; Medicine, Legal; Political Science  &              6 &   352,369 &     14 &       1 \\
\hline

Education         &  Education \& Educational Research; Education, Scientific Disciplines; Education, Special  &              3 &   316,898 &     17 &       2\\
\hline

Building         &  Architecture; Construction \& Building Technology; Engineering, Civil &              3 &   266,346 &     12 &       1 \\
\hline
City Development &  Planning \& Development; Transportation; Transportation Science \& Technology; Urban Studies; regional \& urban planning; development studies &              6 &    86,275 &    19 &       3 \\
	\hline
	\hline
\end{tabular}%
\end{table}%
In the analysis, each article is assigned to exactly one cluster according to the label of disciplines. While 79.26\% of articles has one cluster, including articles with sole subject and articles with multiple subjects belonging to the same cluster, the rest belongs to multiple clusters. Among them, 7.24\% articles are labelled to the cluster with the most common disciplines, and 13.49\% articles with equal amount of subjects belonging to two and more clusters are labelled according to the first discipline in the WoS citation dataset.  

\begin{figure}[ht]
		\includegraphics[width=1\textwidth, height=0.3\textheight]{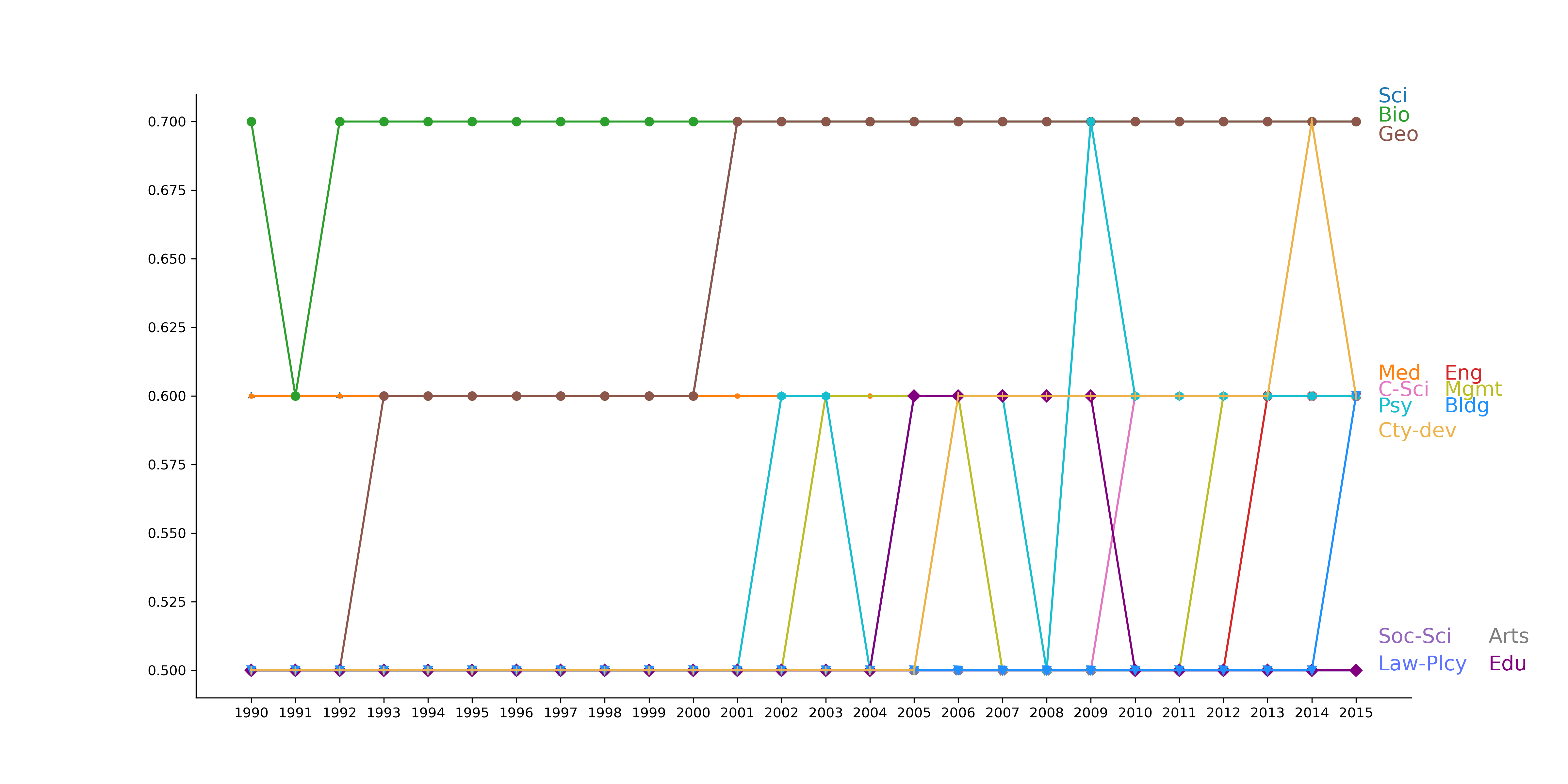}
	\caption{Median ASP for 14 clusters over years between 1990 and 2015.}
	\label{fig:med_asp}
\end{figure}

Figure \ref{fig:med_asp} displays the medians of ASP for each cluster and in each year. Unlike citations that are always monotonically increasing, even only within 5 years citations are considered, ASP provides a different view in terms of scientific prestige of an individual paper. It shows that an article may have both increasing and decreasing prestige values in the citation network, depending on its impact on other articles over time. In general, Biology and Science are more influential with higher ASP than other clusters. Medicine, though with the largest number of papers and citations, has been overtaken by Geography, City Development, Computer Science, Management and Engineering in the recent years, especially after 2000. Law \& Policy, Social Science and Arts have improved their impact in recent years. Nevertheless, their overall impacts are still marginal compared to others. Education and Management are special, with a hump around 2005-2009, which may possibly be their golden age. There is a big drop in Education after that.

\subsection{Hyperparameters choice}
\label{app3}  

\begin{figure}[H]
		\includegraphics[width=0.9\textwidth, height=0.5\textheight]{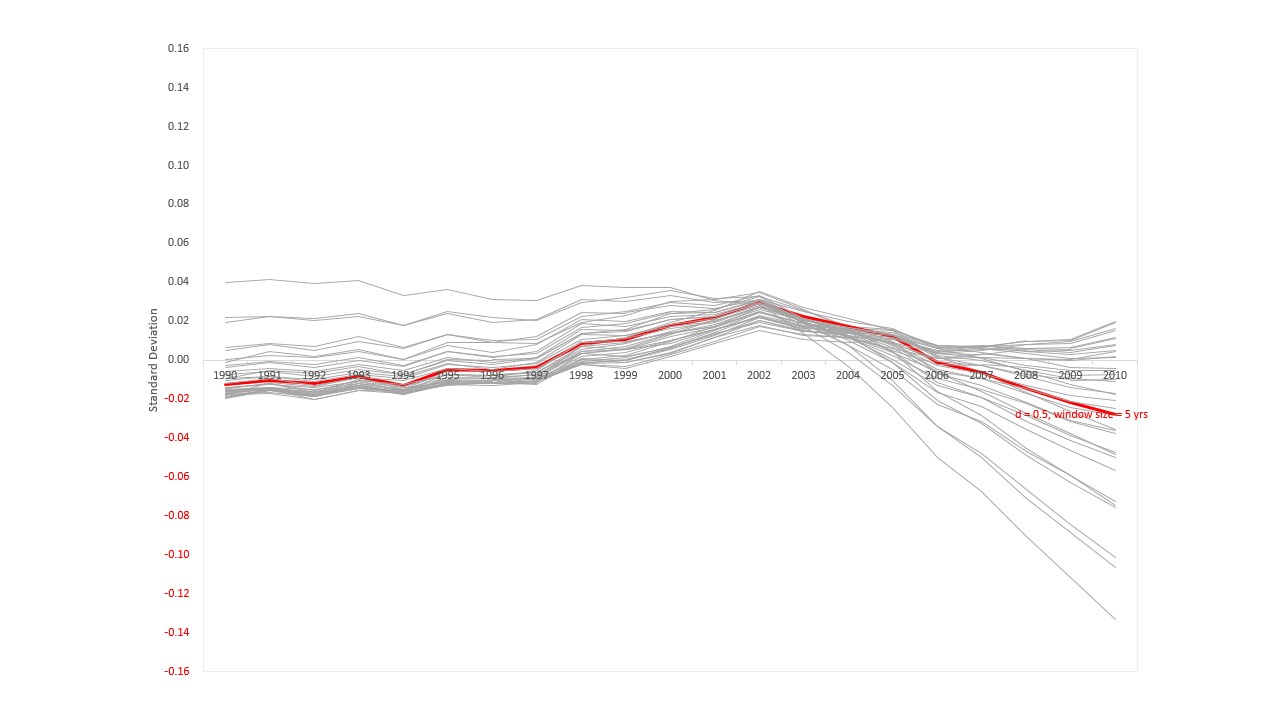}
	\caption{Time evolution of subject scientific impact variations given various combination of damping factor and citing window between 1990 and 2010.}
	\label{fig_stat:dq}
\end{figure}

We conduct sensitivity analysis given different combinations of damping factor $d\in(0.1,0.9) $ and citing window size $\in[1,10]$. To measure the stability, we compare the scaled average value of ASP over years.  Specifically, we compute the average value of ASP in each subject. We display the sum over
the difference between the subject ASP and the average ASP among all articles. To avoid
the time impact, we conduct the computations for each year, see Figure \ref{fig_stat:dq}. It shows that the choice
of $d = 0.5$ and citing window of 5 years led to the minimum deviation among the scientific disciplines. By assuming that no subject is better than another in terms of scientific contribution, we chose the hyperparameters that lead to the minimum variations among
the 254 subjects over years.



\subsection{Articles without any citations}
\label{app4} 
\begin{figure}[H]
		\includegraphics[width=0.9\textwidth, height=0.3\textheight]{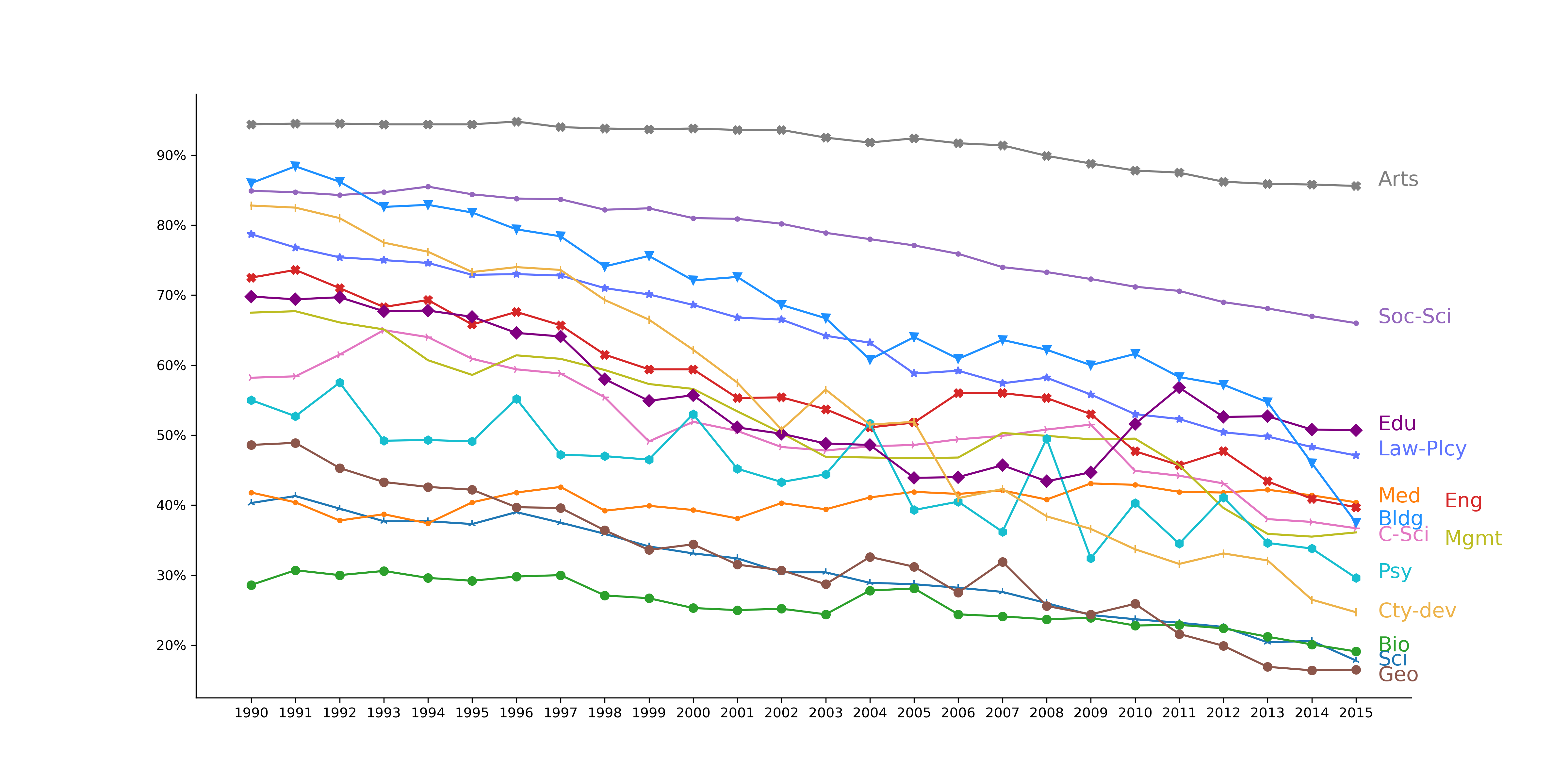}
	\caption{Counting ratio of non-cited articles to total articles per cluster over years between 1990 and 2015.}
	\label{fig_stat:8}
\end{figure}

Figure \ref{fig_stat:8} presents the series of non-cited articles in the 14 clusters over time. Recall that 22.53\% articles have never been cited, the distribution differs among clusters. Science and Medicine have a relatively low ratio of non-citations, i.e. 40\% around 1990-1994 and continuously drops to less than 20\% in 2015. Medicine keeps a stable ratio around 40\%. Geography shows impressive improvement, with the ratio decreasing from 48\% in 1990 to 18\% in 2015. Another cluster City Development reduces the ratio even from 80\% in 1990 to 30\% in 2015. Arts and Social Science have the highest non-cited ratio, where most, e.g. more than 90\% and 78\% articles, are never cited within 5 years. 

\subsection{Correlation of the 14 clusters from 1990 to 2015}
\label{app5}  
\begin{figure}[H]
		\includegraphics[width=0.9\textwidth, height=0.5\textheight]{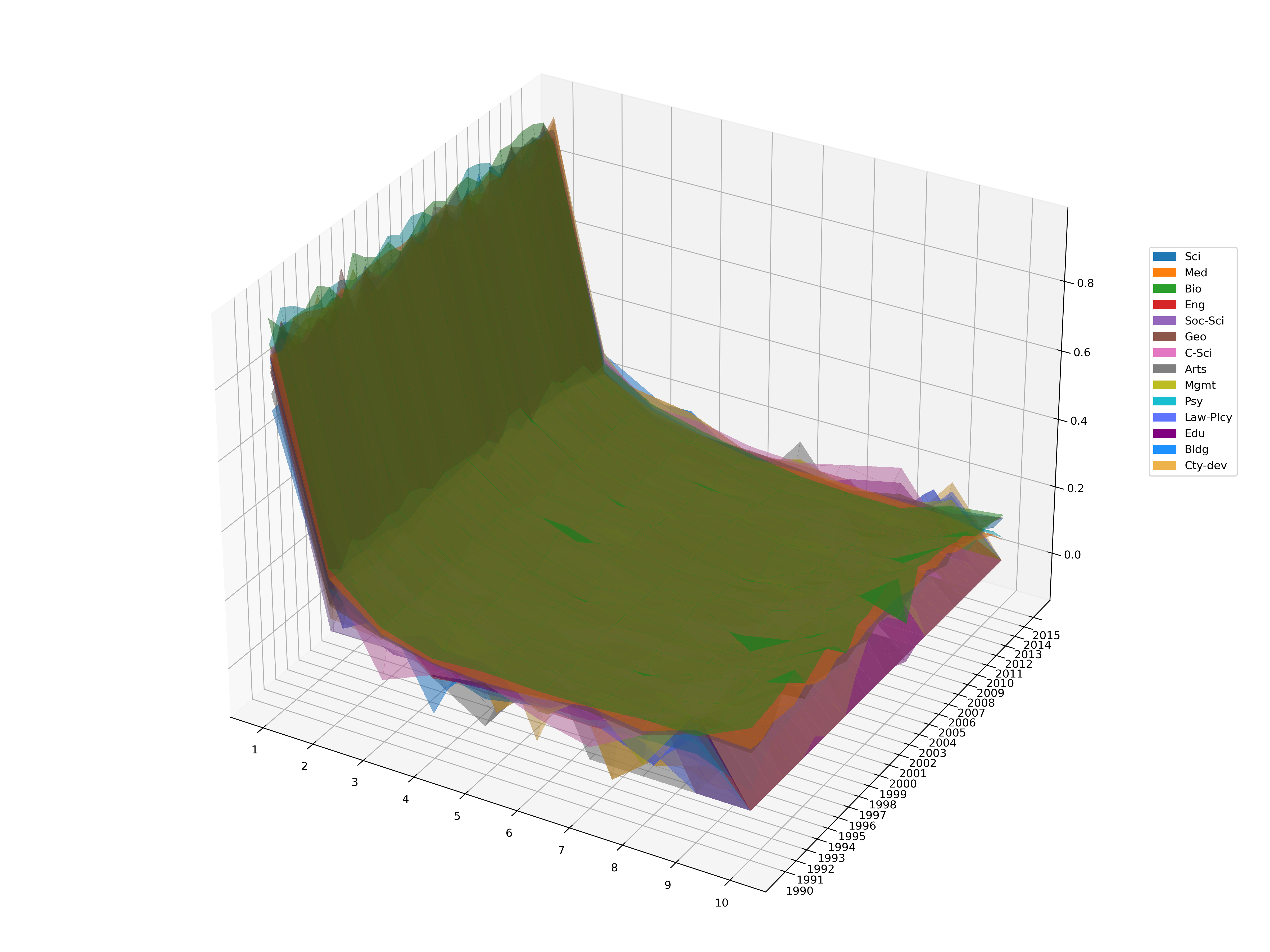}
	\caption{Pearson correlation of ASP and \citcount in 10 groups and over years between 1990 and 2015. Each group includes the correlation values with respect to clusters, deciles and years. Each decile is obtained by dividing articles in each cluster into 10 equal groups according their sorted \citcount. Group 1 belongs to the top 10\% articles with highest \citcount in each cluster. }
	\label{fig_stat:3D}
\end{figure}

Figure \ref{fig_stat:3D} shows the boxplots of Pearson correlation coefficients between ASP and \citcount in 10 decile groups over years from 1990 to 2015. Each group includes $14\times 26$ correlation coefficients with respect to 14 clusters and 26 years. We can see that the correlation is high for the top 10 \% articles with highest \citcount in most clusters. The rest  have lower average correlation (below 0.3). It may indicate that although the top articles have similar high ranks according to their ASP and \citcount, the ranks for the rest  will differ in terms of the two different evaluation metrics. In other words, it is relatively safe to directly use \citcount to evaluate  the highly cited papers (top 10\%), but for the rest, the evaluation of the \citcount may differ significantly from the ASP, which represents the scientific attention of the article.

\end{appendices}

 \newpage










\end{document}